# Results of R&D Programmes and LP/CW E-XFEL Cryomodule Tests in the Period 2005–2023*

J. Sekutowicz[1**], D. Proch[†1], J. Eschke[1], K. Jensch[1], W-D Möller[1], A. Gössel[1], W. Cichalewski[2], A. Bellandi[1], J. Branlard[1], D. Kostin[1], W. Merz[1], N. Mildner[1], F. Mittag[1], R. Onken[1], E. Vogel[1], H. Weise[1], M. Wiencek[1]

[1]*Deutsches Elektronen-Synchrotron DESY, Notkestraße 85, D-22607 Hamburg, Germany*

[2]*TUL, Stefanowskiego 18/22, 90-924 Łódz, Poland*

In 2005 and 2006, we began to consider the feasibility of long-pulse (LP) and continuous-wave (CW) operation for the E-XFEL. The operation modes considered were assumed to be complementary to the short-pulse operation (SP), with ~1 ms RF pulses and a 10 Hz repetition rate, which at that time had already been chosen and presented in the TDR [1] of the E-XFEL facility. This operation mode originated from the previously proposed linear collider TESLA project [2]. We initiated several R&D programmes in 2005 and 2006 to enable operation modes with a duty factor significantly higher than that of the short-pulse mode, which is still approximately 1%. In this report, we briefly present the initiated R&D programmes and their results, with particular emphasis on the results of the E-XFEL cryomodule tests, which led to minor modifications of the design and subsequently to the implementation of these cryomodules in large-scale X-ray FEL facilities.

## I. Introduction

The European XFEL entered operation in 2017, with first lasing achieved in May 2017 and user operation starting in September 2017. The facility operates in SP (short-pulse) mode and delivers up to 2700 bunches during a single RF pulse, separated by a time interval of 222 ns. With a pulse repetition rate of 10 Hz, a total of 27000 bunches per second are delivered to the undulators. In the early years of the R&D programmes, the primary motivation for investigating additional LP (long-pulse) and CW (continuous-wave) operating modes was twofold. First, these modes would allow the delivery of a significantly larger number of bunches to the undulators serving all five E-XFEL photon beamlines. In CW mode, the goal was to deliver at least $2 \cdot 10^5$ bunches per second. Second, they would enable an increase in the bunch spacing from 222 ns to several microseconds, thereby allowing the use of less demanding detector technologies.

The very first parameter set for these operating modes was formulated in 2006 [3], well before the first cryomodule was tested and therefore based on several assumptions made at that time; as a result, it differs from the present parameter set.

Series production of the superconducting RF cavities and cryomodules for the accelerator was scheduled for the period 2012–2015. During the preparatory stage prior to production, two of the initiated programmes were assigned the highest priority due to their impact on the vertical acceptance tests of cavities and the installation of cryomodules in the tunnel:

- Development of high-thermal-conductivity feed-throughs for the HOM (higher-order-mode) couplers, making the cavity end groups less sensitive to heating caused by the fundamental mode and HOMs.
- Development of BLAs (beamline absorbers) to suppress very high-frequency HOMs propagating along the beamline, generated by the large number of electron bunches, particularly when the facility operates in CW mode.

The high-thermal-conductivity feedthroughs were developed in collaboration between DESY and Kyocera Company [4]. Figure 1 shows a photograph of the new HOM feed-throughs developed for LP/CW operation. The feed-throughs are made of high-conductivity materials: pure niobium, molybdenum and sapphire. The niobium antenna is brazed to a molybdenum pin, which is fixed within the sapphire window. This design ensures improved heat removal from the antenna, which is exposed to the magnetic flux of the fundamental mode. Its performance has met expectations in numerous cryogenic tests of cavities and cryomodules.

The E-XFEL linac consists of injector sections comprising 17 cryomodules and the main linac L3, which comprises 80 cryomodules. All 776 accelerating structures of the E-XFEL linear accelerator have been equipped with 1552 feedthroughs of this type [4]. Similar feedthroughs were later implemented in the Linac Coherent Light Source II linac.

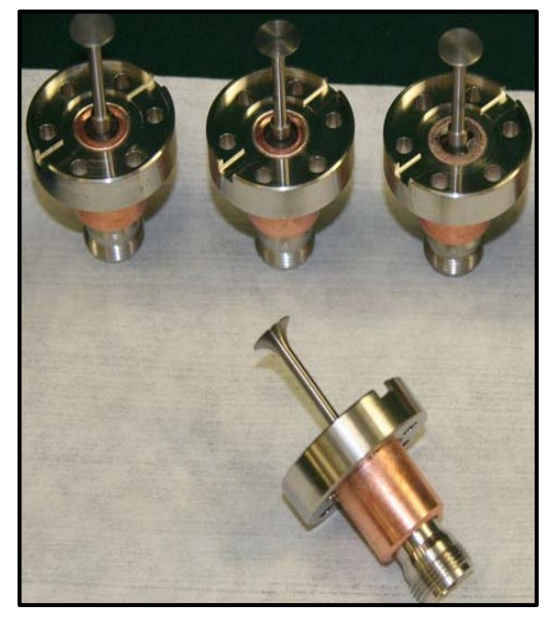
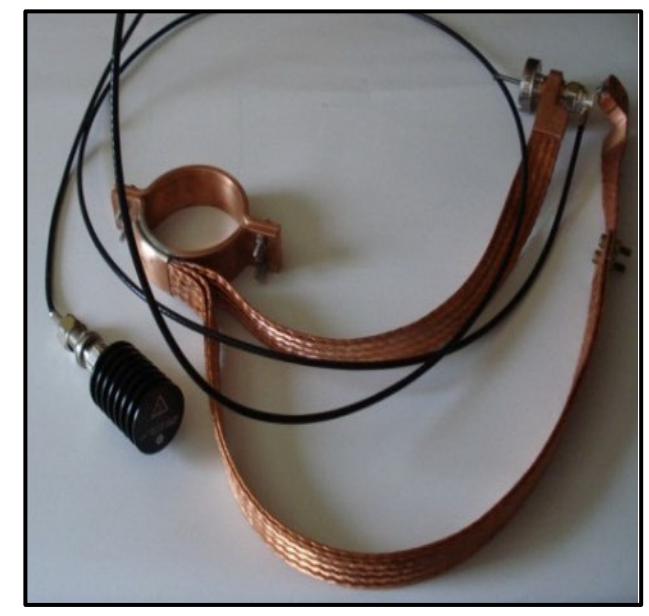

Fig. 1: High thermal conductivity HOM feedthroughs (left) and the thermal connections to the 2 K circuit (right). *Courtesy of NCBJ.*

The BLA, shown in Figure. 2, was developed at DESY [5] and first successfully tested at NCBJ, and subsequently with beam at FLASH in 2008 and 2009 [6]. Tests at NCBJ showed that the BLA can absorb ~100 W of HOM power. More than 160 BLAs were

*Work performed in the framework of R&D for future High Duty Cycle operation modes of the European XFEL accelerator.

**Corresponding author: jacek.sekutowicz@desy.de

[1,2] The authors' affiliations correspond to those held at the time the tests were conducted.

manufactured by Kubara-Lamina S.A. in Poland; 108 are installed at the interconnections between cryomodules in the E- XFEL linac, approximately 30 in the LCLS-II linac, and a further ~30 will be required for the HE (High Energy) upgrade of the LCLS-II linac.

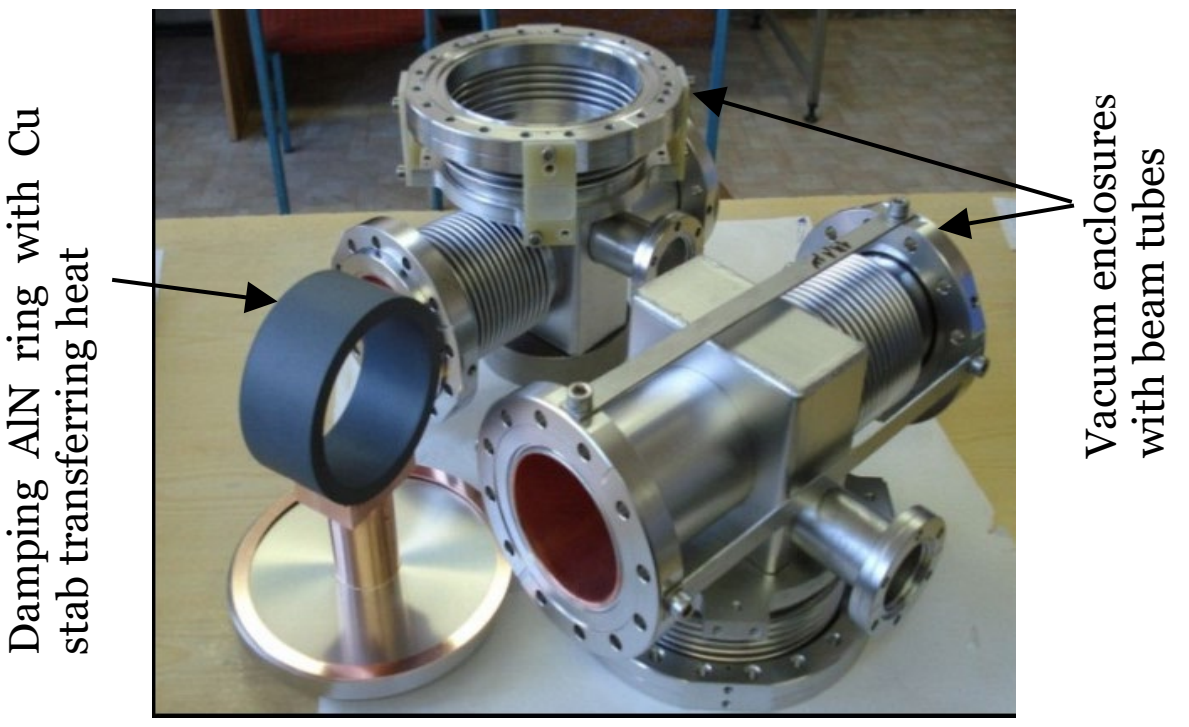

Fig. 2: Parts of the BLA. *Courtesy Kubara-Lamina S.A.*

## II. IOT; RF source for LP/CW operation

New RF power sources will be required for LP/CW operation. The discussion of which type of power source should be chosen for the proposed modes has been ongoing since 2005 and is expected to continue in the coming years.

In 2005–2006, in order to reduce both the investment and operating costs of a new RF system, a high-power Inductive Output Tube (IOT, tetrode) was developed and built by CPI in the USA within the framework of the EuroFEL programme, in which DESY participated from 2005 to 2007. The idea behind this development was to introduce a "modular layout" of the superconducting linac, in which a single RF source supplies power to one cryomodule of the European XFEL type, housing eight TESLA cavities.

The first prototype of the IOT delivered 30 kW of power at 1.3 GHz. Subsequently, CPI agreed to design and manufacture a 120 kW prototype (Figure 3) capable of supplying four cryomodules [7]. This would have enabled the power distribution system to remain the same as that used for high-peak-power klystrons feeding four cryomodules each in SP operation mode. Unfortunately, the second prototype proved to be technically very demanding, and the maximum available power was ultimately about 85 kW (see Table 1). This prototype was installed at the cryomodule test stand, CMTB (Cryomodule Test Bench), and has been available since 2008 for LP/CW tests of European XFEL cryomodules. Over the years, the performance of the second IOT prototype has degraded, and the currently available maximum power is approximately 40 kW.

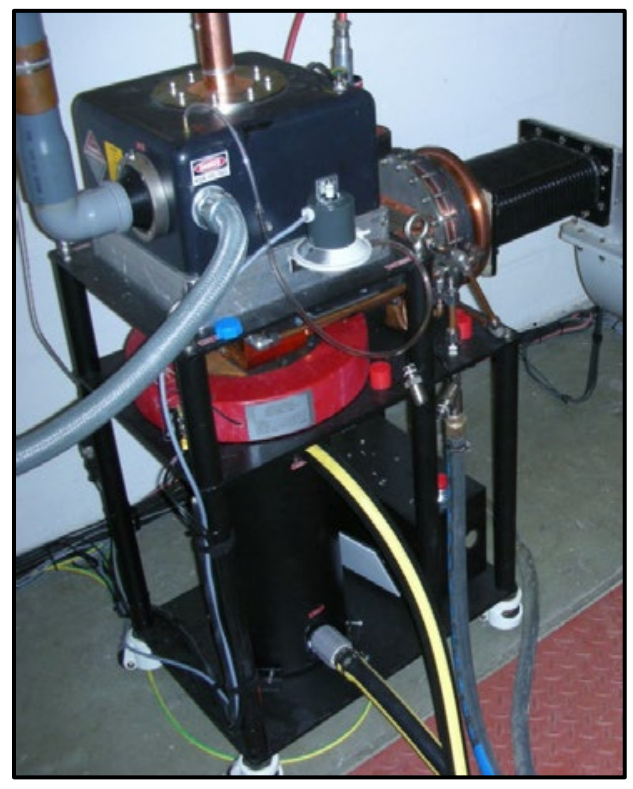

Fig. 3: Second IOT prototype prepared for the test at DESY.

Table 1: Parameters of the second IOT prototype.

| Parameter | Unit | Spec | Measured at CPI / DESY |
|---|---|---|---|
| F | [MHz] | 1300 | 1300 |
| Output P | [kW] | 60 -120 | 85 / 80 |
| Gain | [dB] | >21 | 22.3 |
| Efficiency | [%] | > 60 | 54 |
| Voltage | [kV] | 36 -50 | 45-48 |

Another option could be an integrated IOT station comprising two IOTs delivering 60–70 kW of RF power each. This approach would significantly simplify the design of the IOTs while maintaining the compactness of the RF source, thereby allowing easier installation in the tunnel. An efficiency of about 60% could be expected, with no power consumption between long pulses.

An alternative would be SSAs (solid-state amplifiers), whose efficiency and compactness could improve significantly by the time the ongoing European XFEL HDC (High Duty Cycle) upgrade project is funded.

## III. LP/CW tests of cryomodules at CMTB

We conducted 14 LP/CW tests between 2011 and 2023. Six cryomodules were tested. Five of them housed fine-grain cavities, while one, XM-3, contained seven large-grain cavities and one fine-grain cavity. The XM-3 cryomodule was tested four times: in 2013, 2014, 2017, and 2018.

Two of the tested cryomodules—PXFEL2 (versions PXFEL2.1, PXFEL2.2, PXFEL2.3) and PXFEL3.1—were prototypes. The numbering convention was such that the number following the dot indicated a modification of a cryomodule. This could include, for example, retreatment of cavities, replacement of the fundamental power couplers, repair of a leak, or repair of the tuners.

Two of the tested cryomodules—XM46 (XM46.1) and XM50 (XM50.1)—were subject to minor modifications compared with standard European XFEL cryomodules, whereas the XM4 cryomodule was the standard one.

The first cryomodule tested in LP/CW mode in June-July 2011 was the prototype PXFEL2.1, which had been assembled before the new HOM feedthroughs became available. For this test, eight HOM couplers on the FPC (Fundamental Power Coupler) side (designated HOM1) were equipped with two temperature sensors: one attached to the feedthrough flange and the other at the top of the HOM coupler can (see Figure 4).

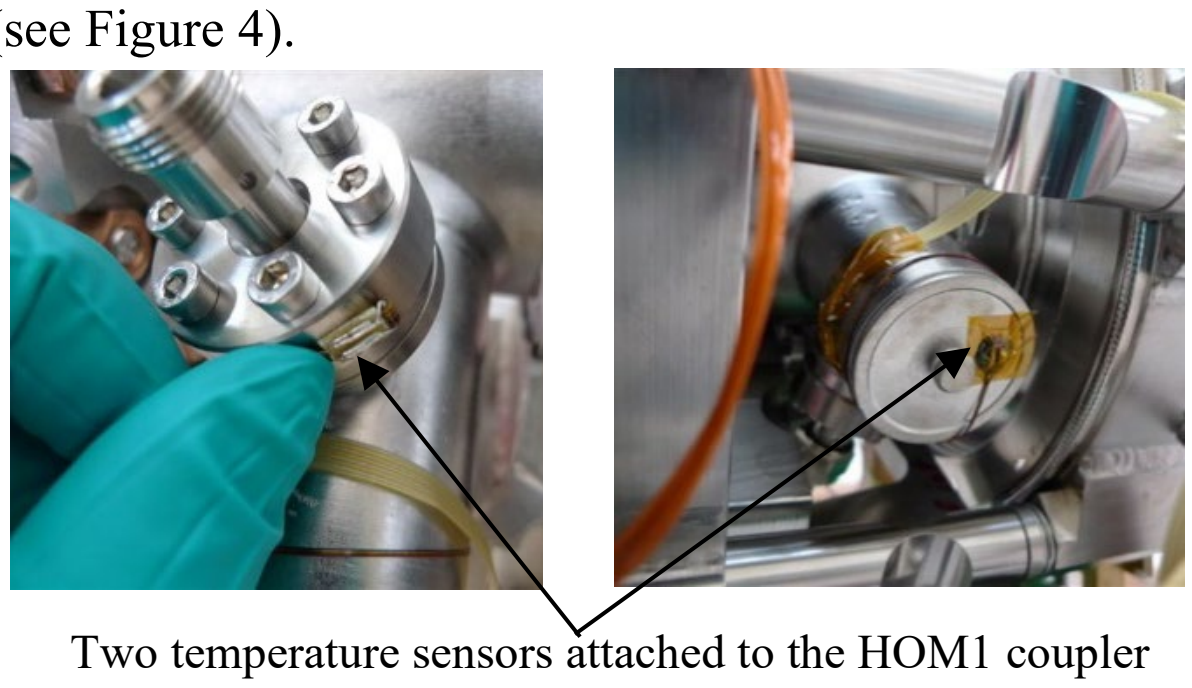

Two temperature sensors attached to the HOM1 coupler

Fig. 4: Temperature sensors attached to the HOM coupler on the FPC side.

The eight couplers on the pickup antenna side (designated HOM2) were equipped with a temperature sensor attached only at the top of the can.

We were extremely cautious during the testing of that cryomodule; therefore, the CW-mode tests were carried out at gradients below 6.5 MV/m while carefully monitoring the temperature of the HOM couplers. Two cavities, numbers 3 and 8, had performed poorly in the previously conducted SP test. Therefore, their FPC $Q_{ext}$ for the LP/CW test was adjusted to $8 \cdot 10^6$ and $3.6 \cdot 10^6$ respectively, while for other six cavities $Q_{ext}$ was set to ~$1.6 \cdot 10^7$.

Figure 5 shows the gradient of all cavities during a 1-hour CW run. The gradients of cavities #3 and #8 were lower due to their lower $Q_{ext}$. Cavity #4 also demonstrated a lower gradient, but this was due to an imperfectly balanced RF-power distribution system, which resulted in less power being delivered to that cavity. Figure 6 shows the gradient and out-coupled HOM power of cavity #6. While the power of HOM2 stays constant, the power of HOM1 increases significantly over time, indicating warming and thus detuning of the fundamental mode filter in that coupler. Figure 7 shows the temperature at the flange of the HOM1 coupler of cavity #6 during the test period.

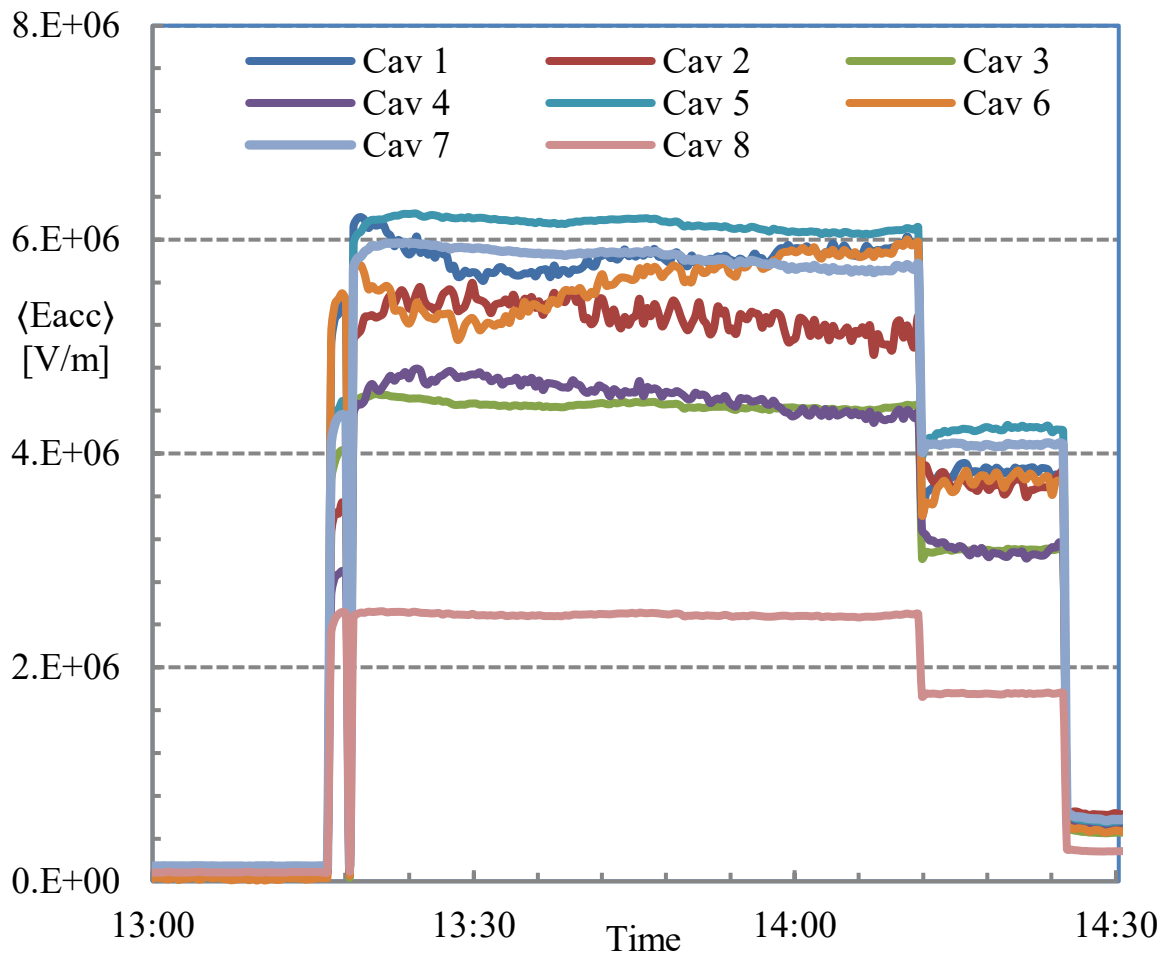

Fig. 5: Gradients of 8 cavities during CW run.

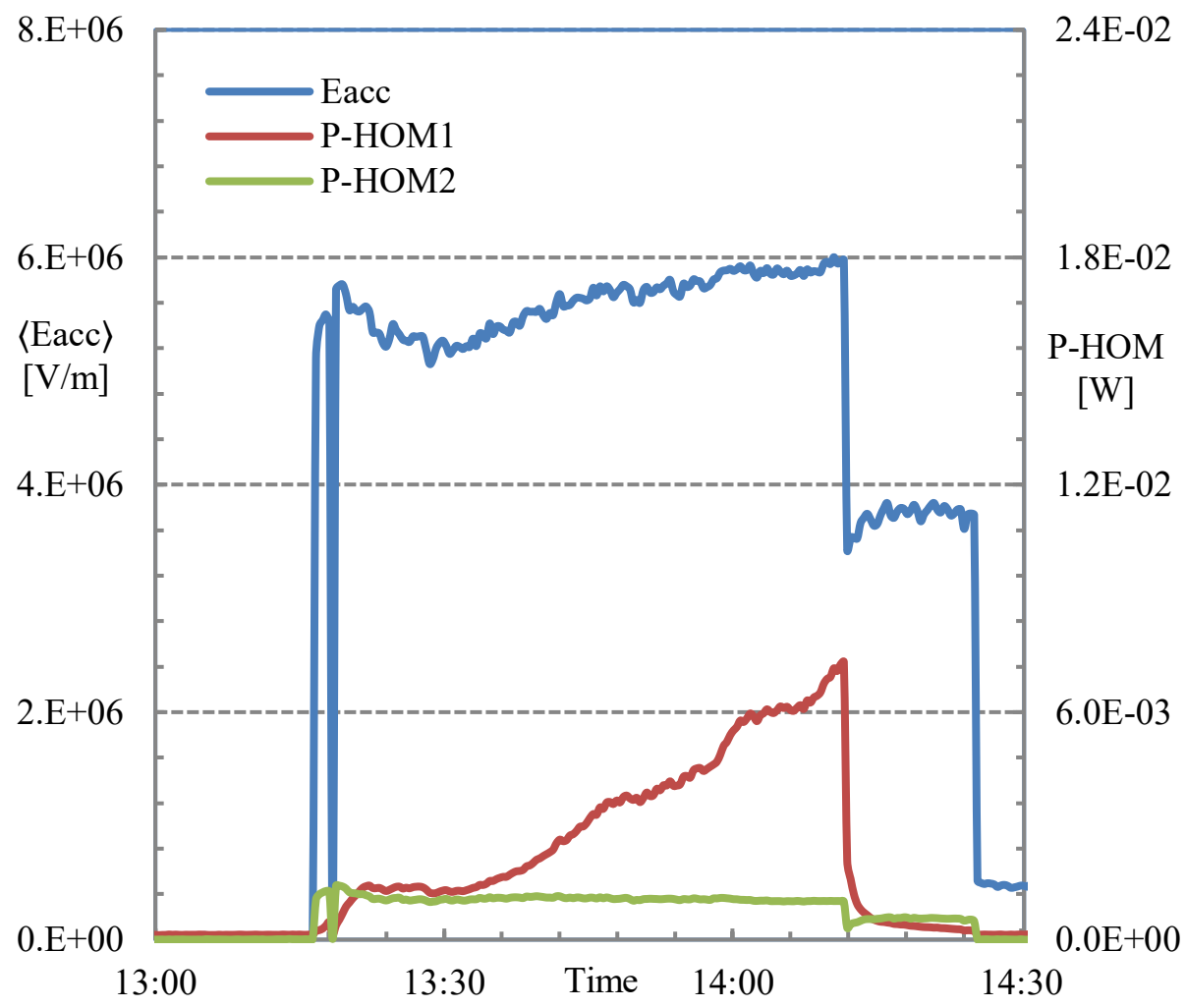

Fig. 6: Gradient and out-coupled HOM power of cavity #6.

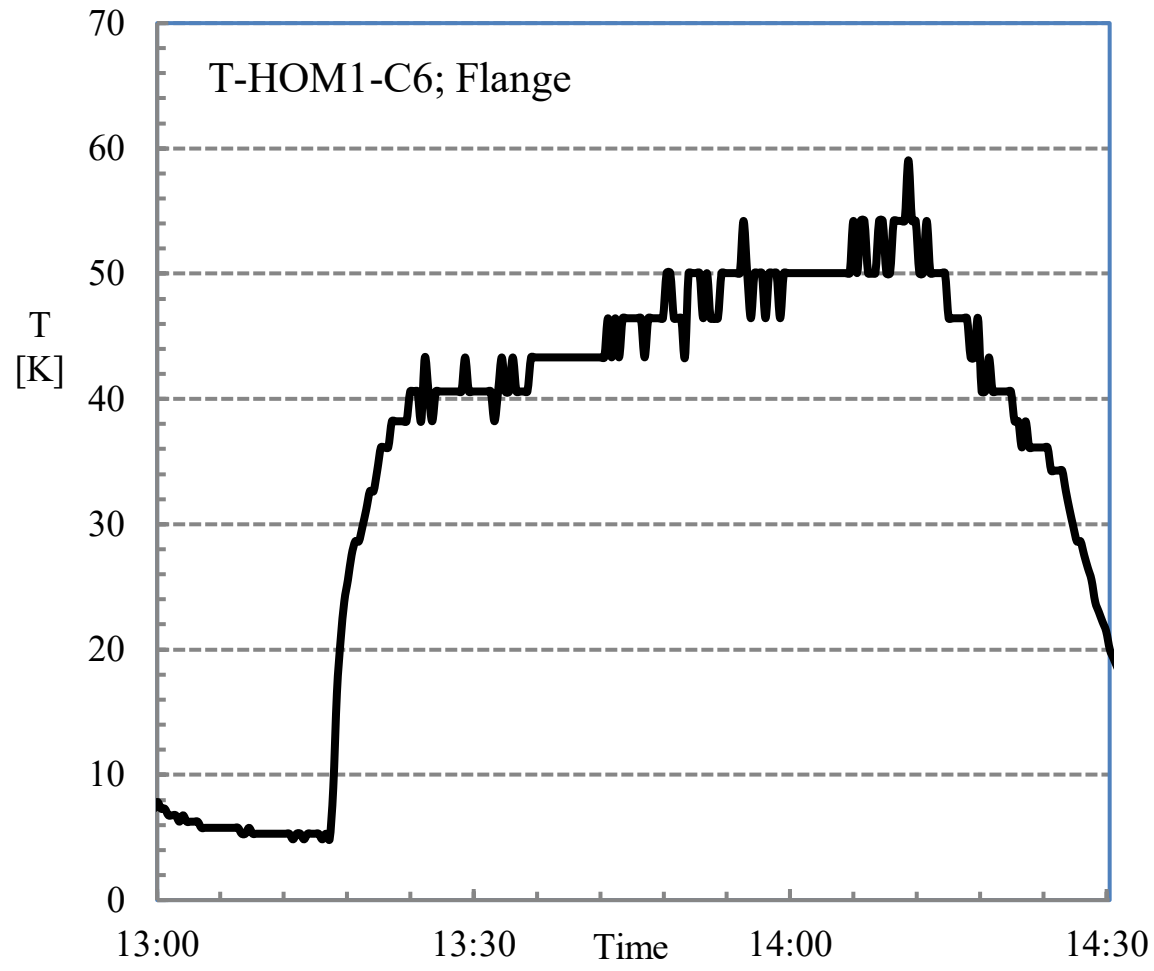

Fig. 7: Temperature at the flange of HOM1 coupler of cavity #6.

Based on this observation, we concluded that during future LP/CW tests any substantial change in the out-coupled power from a HOM coupler would indicate warming of that coupler. This eliminated the need for expensive temperature sensors, whose installation was also a time-consuming step during cryomodule assembly.

In 2012 we conducted tests of PXFEL2.2 and PXFEL3.1 [8]. The tests focused on measuring the cryogenic heat load at 2 K as a function of gradient and duty cycle. The aim was to determine at what gradient the L3 linac of the EXFEL accelerator could operate in LP/CW modes. L3 is the longest linac in the accelerator, and a further 16 cryomodules could be added to its 80 cryomodules for the LP/CW upgrade. The cryogenic system supplies each cryomodule with approximately 1 g/s of superfluid helium to remove a heat load of 20 W at 2 K. For cost reasons, the proposed LP/CW upgrade requires L3 to remain unchanged for SP mode operation, and the only possible modifications are the relocation of 16 existing cryomodules from the injector section of the accelerator to the end of L3, and an increase in the $Q_{ext}$ of the FPCs. Figure 8 shows the measured total heat load (dynamic plus static) at $\langle E_{acc}\rangle$= 5.6 MV/m from tests conducted in 2012 on the prototype cryomodules PXFEL2.2 and PXFEL3.1, as well as the maximum $\langle E_{acc}\rangle$ as a function of duty factor for the limit of ~20 W heat load at 2 K per cryomodule.

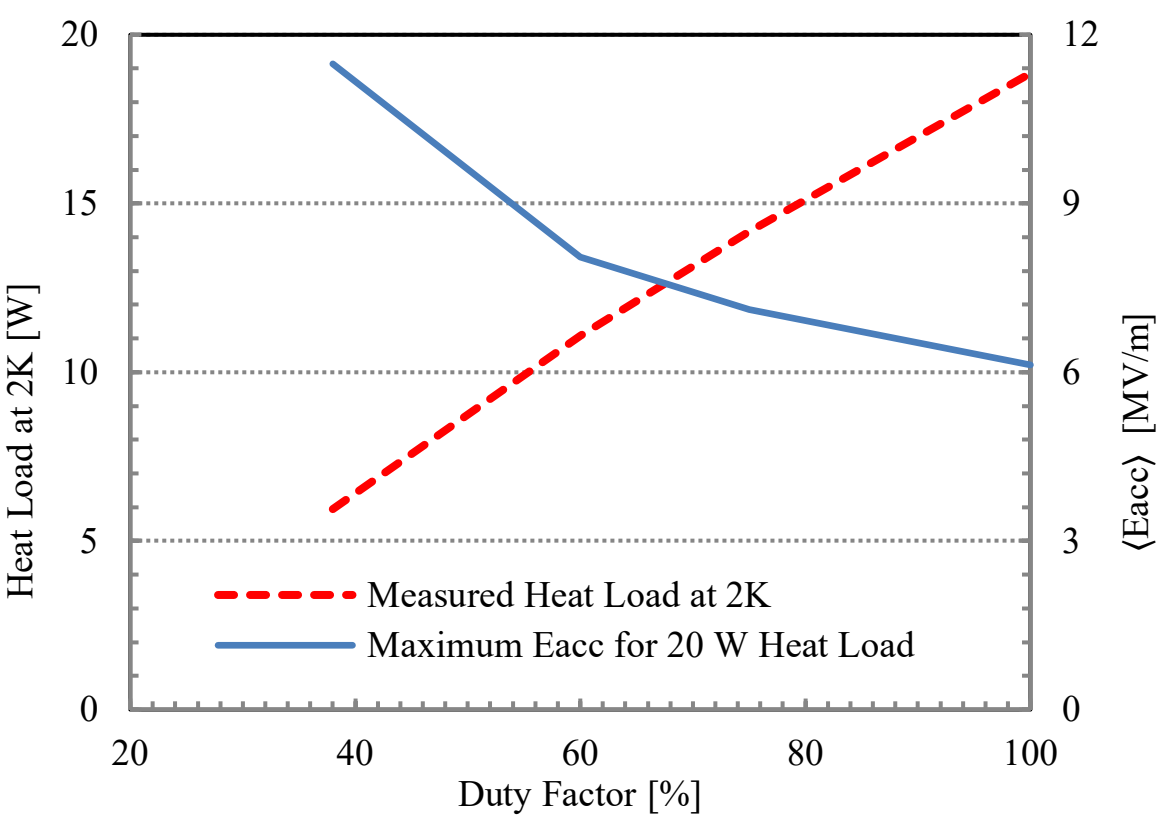

Fig. 8: Measured heat load from tests conducted in 2012 on the prototype cryomodules PXFEL2.2 and PXFEL3.1, and the maximum $\langle E_{acc}\rangle$ as a function of the duty factor.

Cryomodule PXFEL3.1 was the first in which the HOM couplers of five cavities were equipped with high-conductivity feedthroughs—four manufactured by the U.S. company Omley Industries and one by the Japanese company Kyocera—which were subsequently adopted for all series cryomodules. Figure 9 shows the way in which the feedthroughs were thermally connected to the two-phase tube by means of a copper braid. Figure 10 presents the dynamic heat load (DHL) measured over a 24-hour period during a long-term run of cryomodule PXFEL3.1 at $\langle E_{acc}\rangle$ =5.6 MV/m.

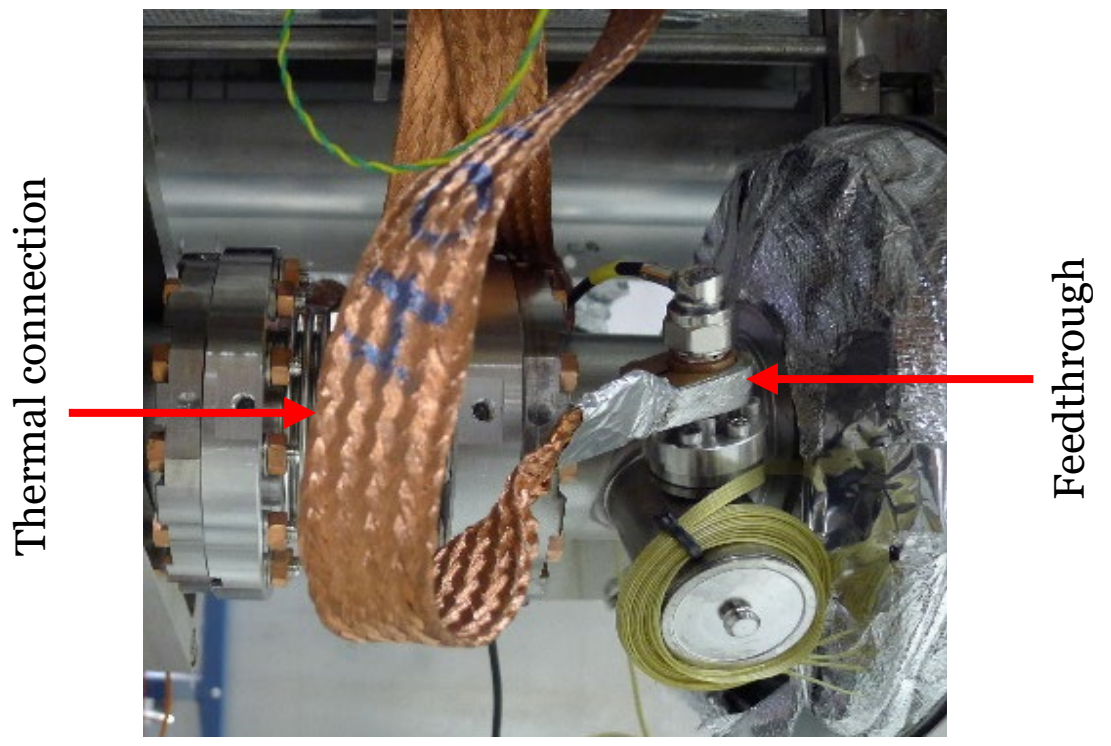

Fig. 9: Thermal connection with 2-phase tube of the HOM coupler new feedthrough by Kyocera.

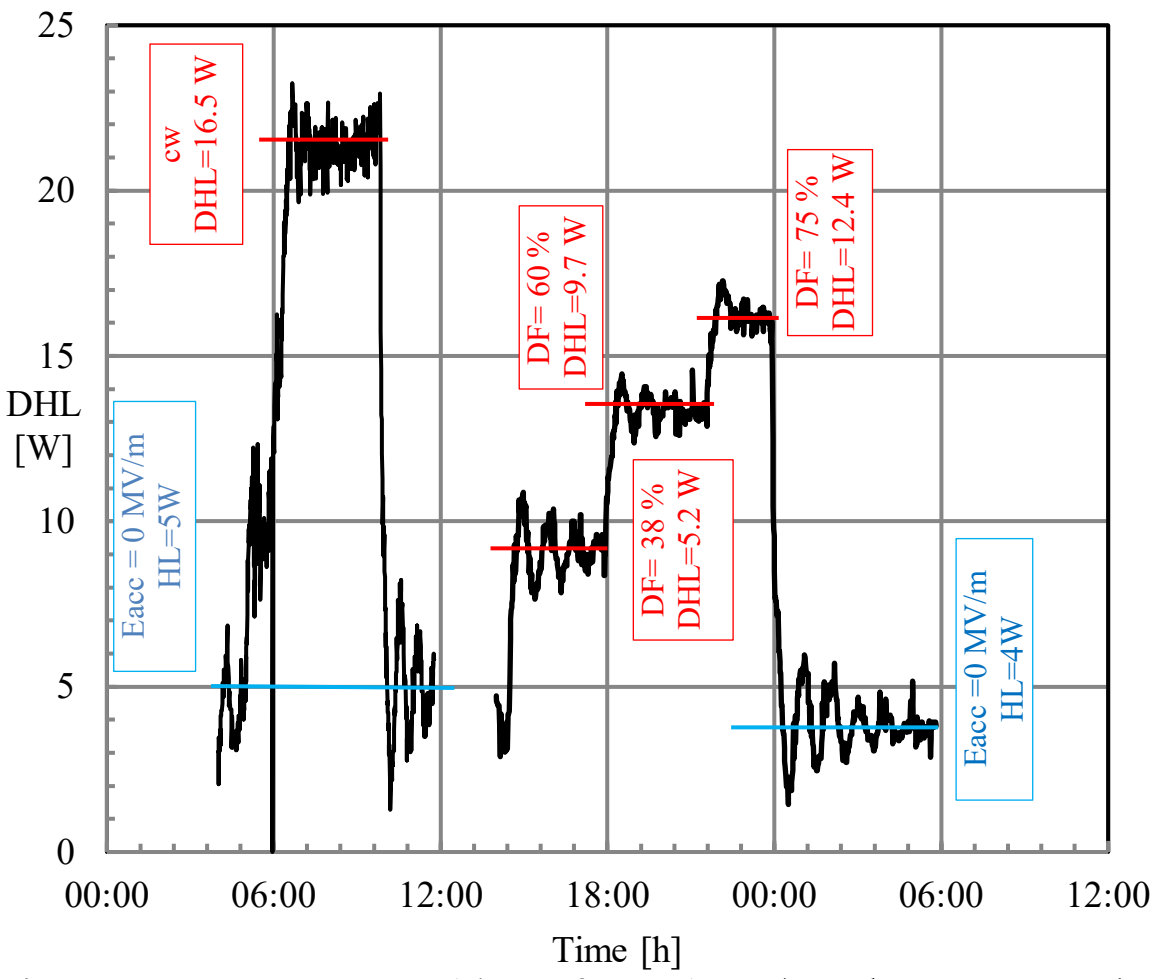

Fig. 10: DHL vs. DF (duty factor) at $\langle E_{acc}\rangle$ = 5.6 MV/m and T=2K. Reference levels at $E_{acc}$= 0 MV/m, are marked in blue.

In parallel with the heat load tests, an intensive R&D programme on LLRF was continued in 2012, mainly using cryomodule PXFEL3.1. The main achievements were as follows:

- Testing of a new hardware architecture for LLRF (μTCA)
- Testing of LLRF controllers (firmware, RF–piezo)
- Testing of DOOCS (Distributed Object-Oriented Control System) servers (software, RF–piezo)
- Characterisation of cryomodule PXFEL3.1.

In the subsequent years up to the present, the LLRF R&D programme has been continuously pursued, with the objective of achieving, for LP/CW operation modes, stability of the accelerating gradient amplitude and phase at the levels of 0.01% and 0.01°, respectively, as presently demonstrated for the SP operation mode [9].

In April 2013, testing of the next cryomodule, PXFEL2.3 (the third version of PXFEL2), began. Figure 11, which shows DHL versus time, summarises a 13-hour test of the PXFEL2.3 cryomodule at various gradients, with the DF held constant at 20%. The gradients varied from 7.4 MV/m to 10.7 MV/m, and for each gradient the testing time was at least 45 minutes. We did not observe any abnormal behaviour of the cryomodule during the test period. The end groups of all eight cavities were thermally stable, and the out-coupled fundamental mode (FM) power via the HOM couplers was proportional to the input power, indicating that the HOM coupler filters for the FM did not heat up [10].

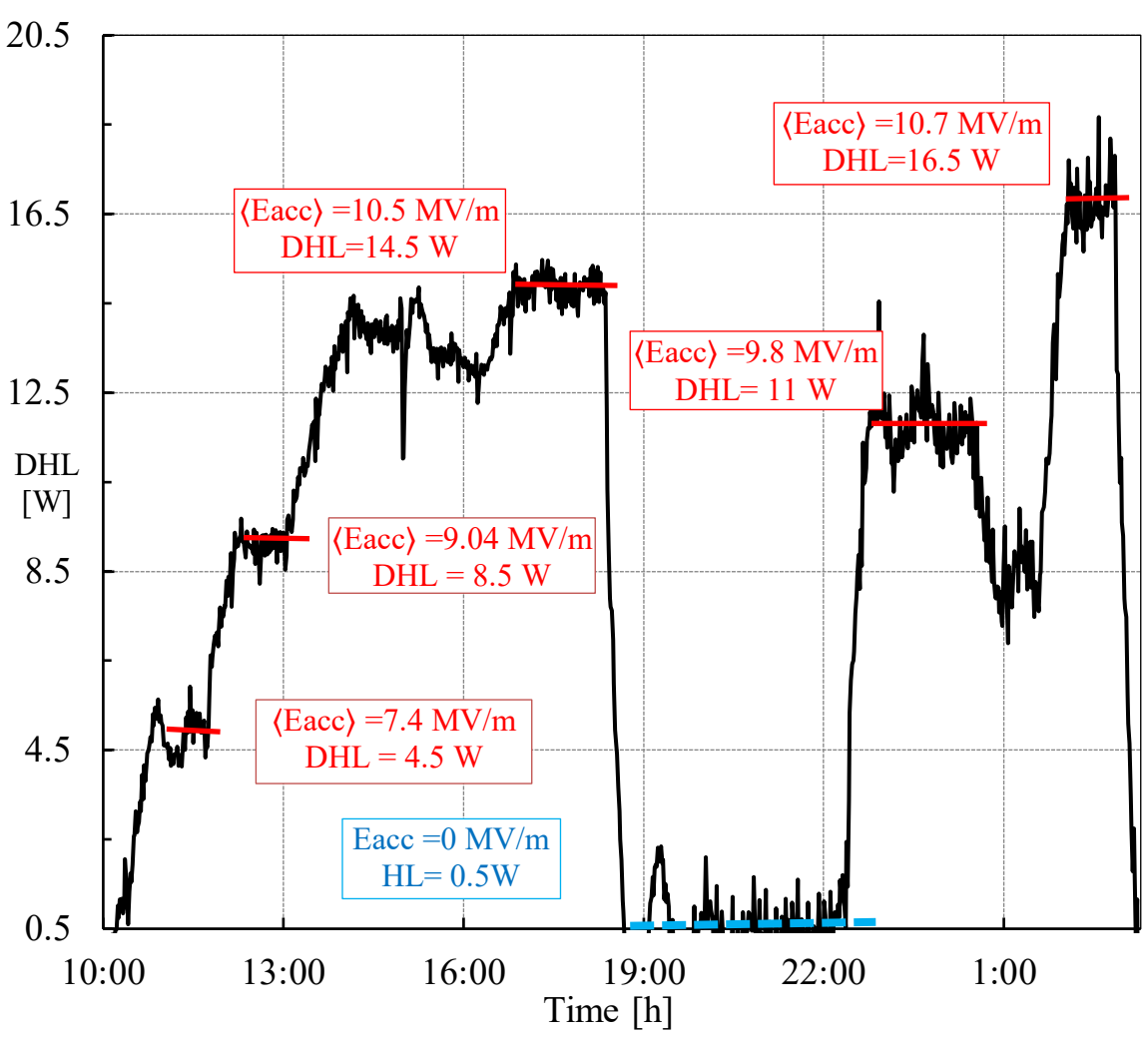

Fig. 11: PXFEL2.3; DHL vs. $\langle E_{acc} \rangle$ for DF = 20%. Reference level at $E_{acc}$ = 0 MV/m, marked in blue, is heat load HL= 0.5 W.

In the period from September 2013 to January 2014, we focused on investigating the impact that the speed of transitions through the critical temperature of niobium ($T_c$ = 9.2 K) might have on the intrinsic $Q_0$ of the cavities. To this end, we tested three cooldown procedures on cavities housed in the XM-3 cryomodule. As previously mentioned, XM-3 housed seven large-grain cavities and one fabricated from fine-grain niobium. All HOM couplers of the eight cavities were equipped with new feedthroughs, thermally connected the two-phase tube.

The transition speed through Tc in the first procedure was approximately −3 K/min. In the second procedure, the speed was very low, approximately −0.01 K/min. The third procedure consisted of two steps. First, the cavities were cooled down to 15 K and held at this temperature for 12 hours. Subsequently, they were cooled down to approximately 4 K at a high rate of about −4 K/min. The differences in the mean intrinsic $\langle Q_0 \rangle$ of the eight cavities at 1.8 K and 2 K, measured in CW mode at $\langle E_{acc} \rangle$ = 7 MV/m, are summarised in Table 2. As expected, the slow cooldown (second procedure) resulted in a $\langle Q_0 \rangle$ that was lower by 30% and 20% at 1.8 K and 2 K, respectively. By contrast, the difference in $\langle Q_0 \rangle$ between the moderate and very fast cooldown is marginal and lies within the measurement uncertainty [11].

Table 2: Intrinsic $\langle Q_0 \rangle$ at 1.8 K and 2 K.

| Date | Trans. speed [K/min] | <Qo> at 1.8 K [$10^{10}$] | <Qo> at 2 K [$10^{10}$] |
|---|---|---|---|
| 12-13.09.2013 | -3 | 4.7 | 3.1 |
| 13.12.2013 | -0.01 | 3.3 | 2.5 |
| 25-27.01.2014 | -4 | 4.7 | 3.2 |

In 2015, during the period from June to December, we tested cryomodule XM4. Two main observations were made that had an impact on the modifications required for 17 cryomodules in the injector section, responsible for generating and forming electron bunches (linacs L1 and L2), in the context of the LP/CW upgrade.

First, careful tuning of the HOM couplers' filters, rejecting the fundamental mode, is crucial. High outcoupled power leads to warming of a HOM coupler, which in turn causes further detuning of the filter. The filter of the HOM coupler attached to the beam tube on the FPC side (HOM1) of cavity #2 was found to be detuned. The $Q_{ext}$ of that coupler was $1.4 \cdot 10^{10}$. The outcoupled power reached 90 W at the maximum gradient of 35 MV/m achieved by this cavity in the SP mode test. The average outcoupled power was only 0.8 W due to the very low duty factor (DF = 0.88%), and no warming of the coupler was observed during the SP test. However, during the CW test at gradients ≥ 10 MV/m, warming of the coupler and end group was observed when the run duration exceeded approximately 30 minutes. The warming of the HOM1 coupler and the end group also led to an increase in liquid helium consumption by 0.3 g/s in the 2 K circuit, indicating that power dissipation in that location increased by approximately 6 W.

Secondly, during CW operation, we observed significant warming of the FPCs, causing a reduction in the loaded quality factors ($Q_{load}$) of the cavities. All FPCs in all cryomodules are equipped with two PT1000 sensors (for redundancy) attached to the cold side beneath the 70 K flange connecting the cold and warm parts of the inner conductor, allowing indirect monitoring of the temperature rise of the inner conductors. The cold part of the inner conductor is made of high-purity bulk copper; therefore, a relatively small temperature gradient along the cold part is expected. Figure 12 shows $Q_{load}$ as a function of the flange temperature indicated by the sensor.

The duration of this CW test was 4 hours. The average gradient of the seven cavities was 14 MV/m, while cavity 2 was detuned and operated at 2.5 MV/m. The mean loaded quality factor, $\langle Q_{load} \rangle$, decreased by 20% during the run. The output power of the IOT ($P_{IOT}$) was monitored and regulated by the LLRF system to maintain constant gradients throughout the test. At the beginning of the test, $P_{IOT}$ was 28.2 kW and increased to 37.4 kW by the end of the run, corresponding to a 25% increase in power. The additional 5% can be attributed to increased dissipation in the RF power distribution system and/or to the accuracy of the measurements. To first order, the product of the loaded quality factor and the power delivered to the cryomodule remained constant, which is consistent with unchanged gradients during the run.
The FPC warming phenomenon forms a positive feedback loop and does not stabilise on its own. A lower loaded quality factor requires higher power to maintain stable gradients, and the increased power leads to greater warming of the coupler. Consequently, a decision has been made for the FPCs in the new cryomodules in L1 and L2 to increase the thickness of the warm part of the inner conductor copper coating from the current 30 µm to approximately 150 µm. This will improve thermal conductivity and help remove heat from the inner conductors.

One further achievement realised during the XM4 test period is also worth noting. The LLRF group at DESY implemented Active Noise Compensation (ANC), using piezoelectric stacks for microphonics suppression, integrated into the LLRF system for both

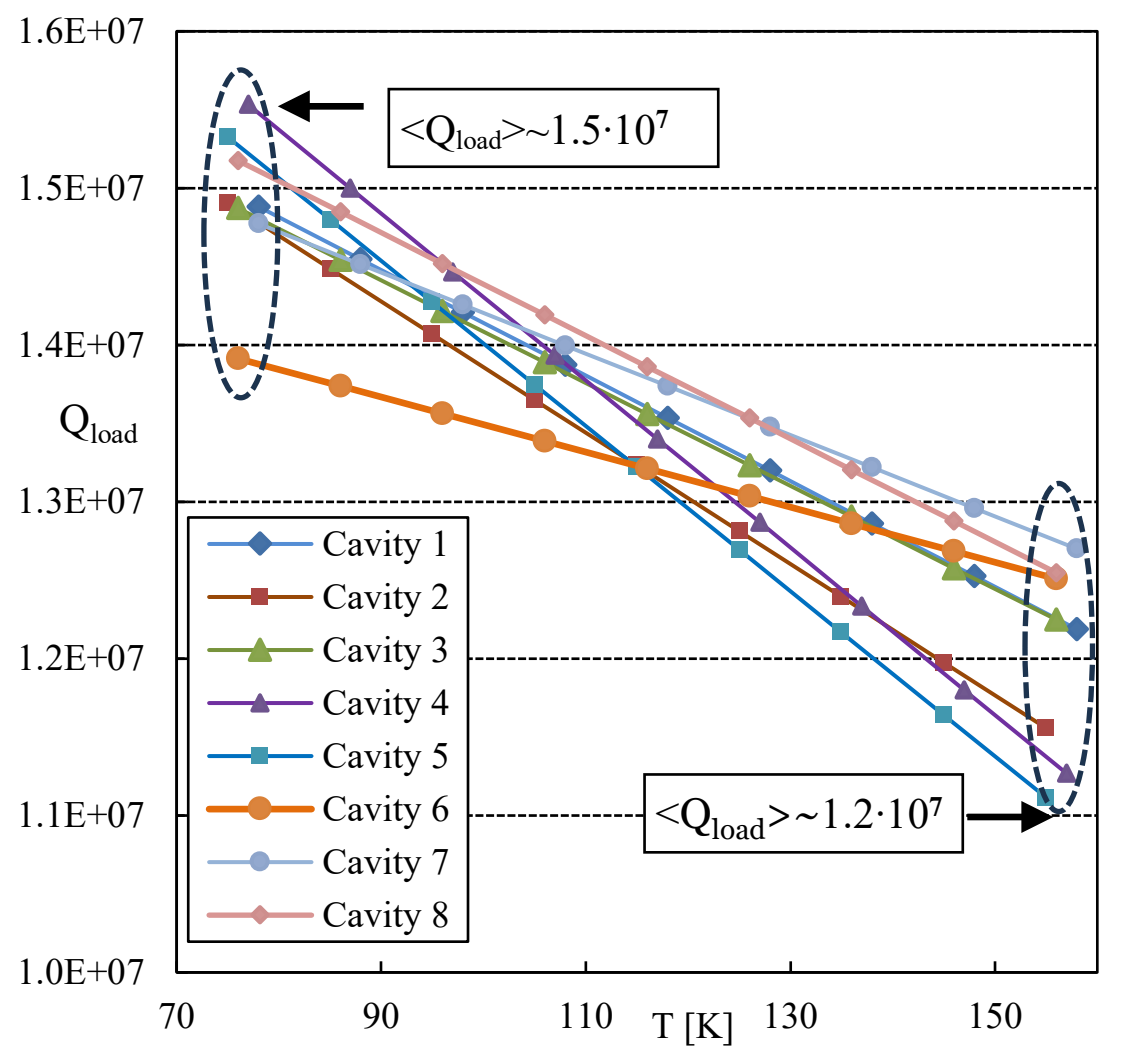

Fig. 12: Change of $Q_{load}$ of the FPCs vs. temperature measured by the PT1000 sensor.

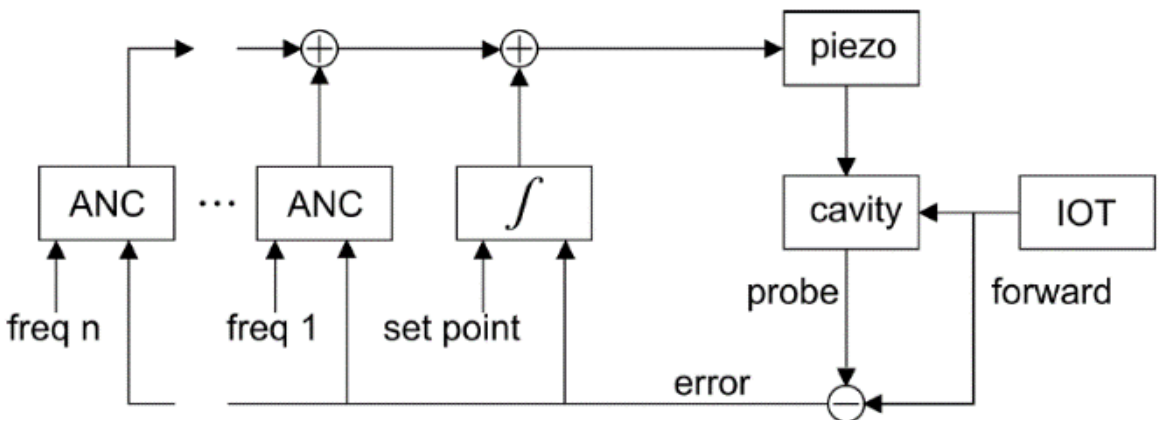

Fig. 13: Piezo fast tuner with ANC and integral feedback controller.

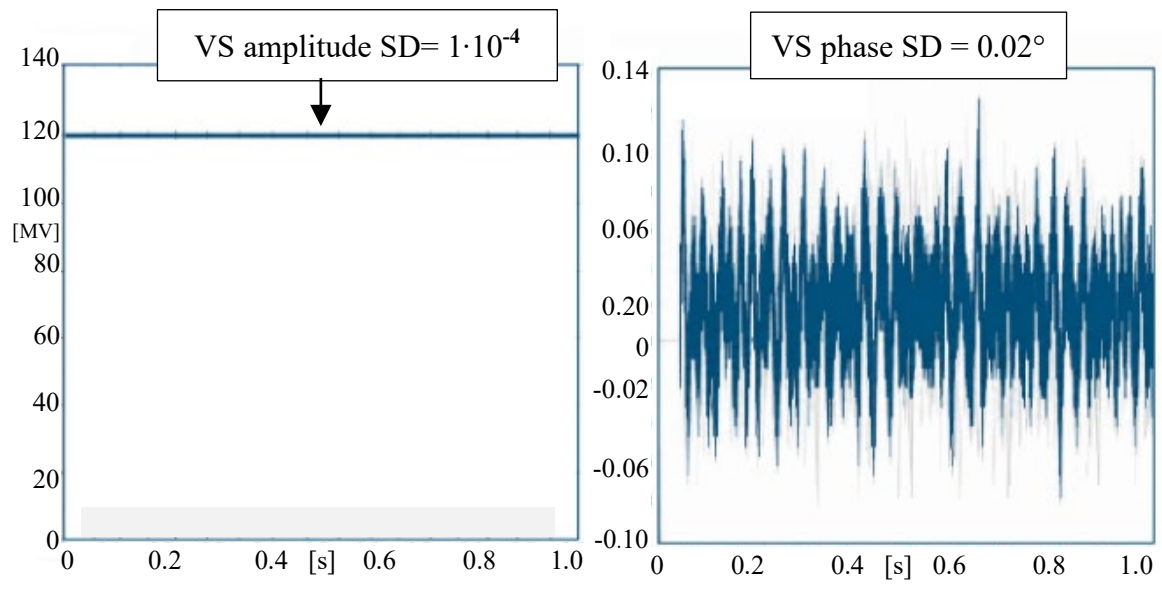

Fig. 14: Example of the suppression of microphonics for operation at 15 MV/m in CW mode.

LP and CW operation. Figure 13 shows the block diagram of the ANC system. Figure 14 displays Vector Sum (VS) standard deviation (SD) for amplitude (left) and phase (right). The amplitude stability fulfils specification for the SP mode operation, while the phase stability falls short of the specification by a factor of two.
The measured dependence of the amplitude standard deviation on the gradient is shown in Figure 15. Stability improves as the gradient increases. This effect may be due to the simultaneous deformation of the cavity wall by a constant microphonics force and the significantly larger Lorentz force, which is proportional to the square of the gradient, $\langle E_{acc}\rangle^2$.

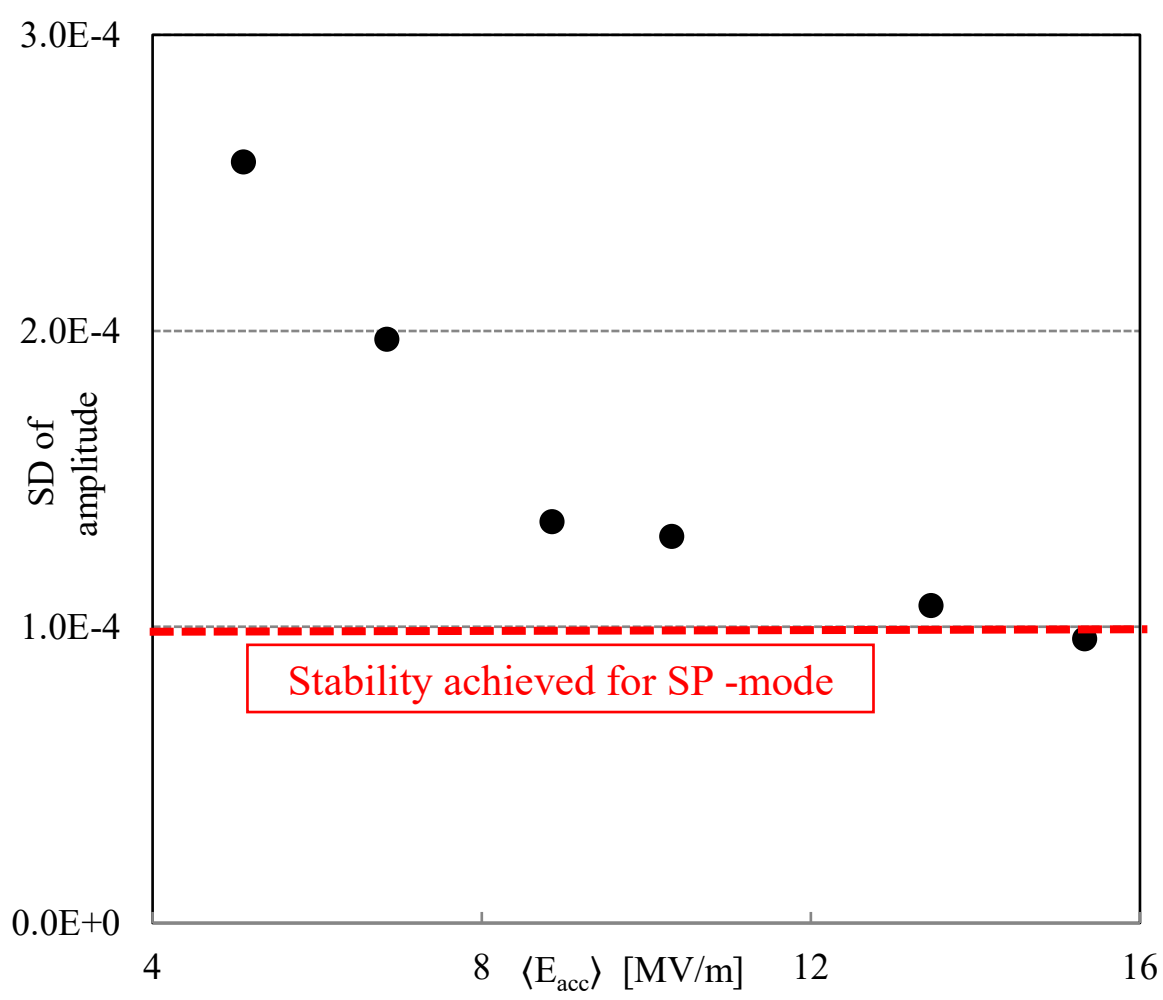

Fig. 15: Dependence of the amplitude SD as function of gradient.

In October 2015, the XM46 cryomodule was pre-tested in SP mode. The pre-test showed that six of the eight cavities were limited by X-ray emission below the acceptance specification for gradient. In addition, the HOM2 coupler on the pickup side of cavity 1 had an improperly tuned filter, and its $Q_{ext}$ was $1.4{\cdot}10^{10}$.
In April 2016, the cryomodule was installed at CMTB for further investigation. The initial focus was on the dark current (DC) that the cryomodule might generate. For this purpose, two Faraday cups were attached to the cryomodule, one on each side.
First, all cavities were tested individually in CW mode. In each test, only the cavity under investigation was tuned to resonance, and the gradient was increased stepwise, while the dark current on both sides and the X-ray emission were monitored. These measurements showed that cavity No. 6 emitted the highest dark current in the direction of cavity No. 1. Cavity No. 6 also produced the highest level of X-ray radiation.
Then, the gradient of cavity No. 6 was set to 10.5 MV/m. Subsequently, proceeding cavity by cavity in sequence from cavity No. 5 to cavity No. 1, the gradients and phases of each cavity were adjusted to maximise the dark current in the Faraday cup on that side. This procedure resulted in a very high dark current of 505 nA and radiation level of 176 mGy/min, whereas the limit for X-ray radiation in SP operation mode is 0.1 mGy/min. Seven cavities of this cryomodule were subsequently re-treated and demonstrated high gradients in vertical tests. One cavity was replaced with a new unit that performed better in the vertical test. The mean operating gradient of the new version, XM46.1, when all cavities meet the specifications, is 33 MV/m. The LP/CW operation of this cryomodule is discussed in more detail later in this publication.

During the period from 17.07.2019 to 21.01.2020, we tested cryomodule XM50.
In period from 17.07.2019 to 16.09.2019 and on 20.10.2019 FPCs, had $Q_{ext}$ adjusted to ~$3{\cdot}10^7$. We measured $Q_o$ as a function of $\langle E_{acc}\rangle$ up to approximately 10 MV/m for all eight cavities. Beyond this gradient, cavity No. 2 exhibited very strong radiation, and testing was continued with seven cavities. Unfortunately, in July 2019, the IOT was out of service for two months, and for one week the cryogenic system experienced issues, resulting in the cryomodule being held at 40 K and then rapidly cooled back down to 2 K. The $Q_o$ for 5.7 MV/m and 8.2 MV/m was unexpectedly high. This can be attributed to the low accuracy of the DHL data when the dynamic losses are small.

Monitoring of the outcoupled fundamental mode power by the HOM couplers showed that, in all these runs, the filters were not warmed up; the measured power remained proportional to the stored energy in the cavities.

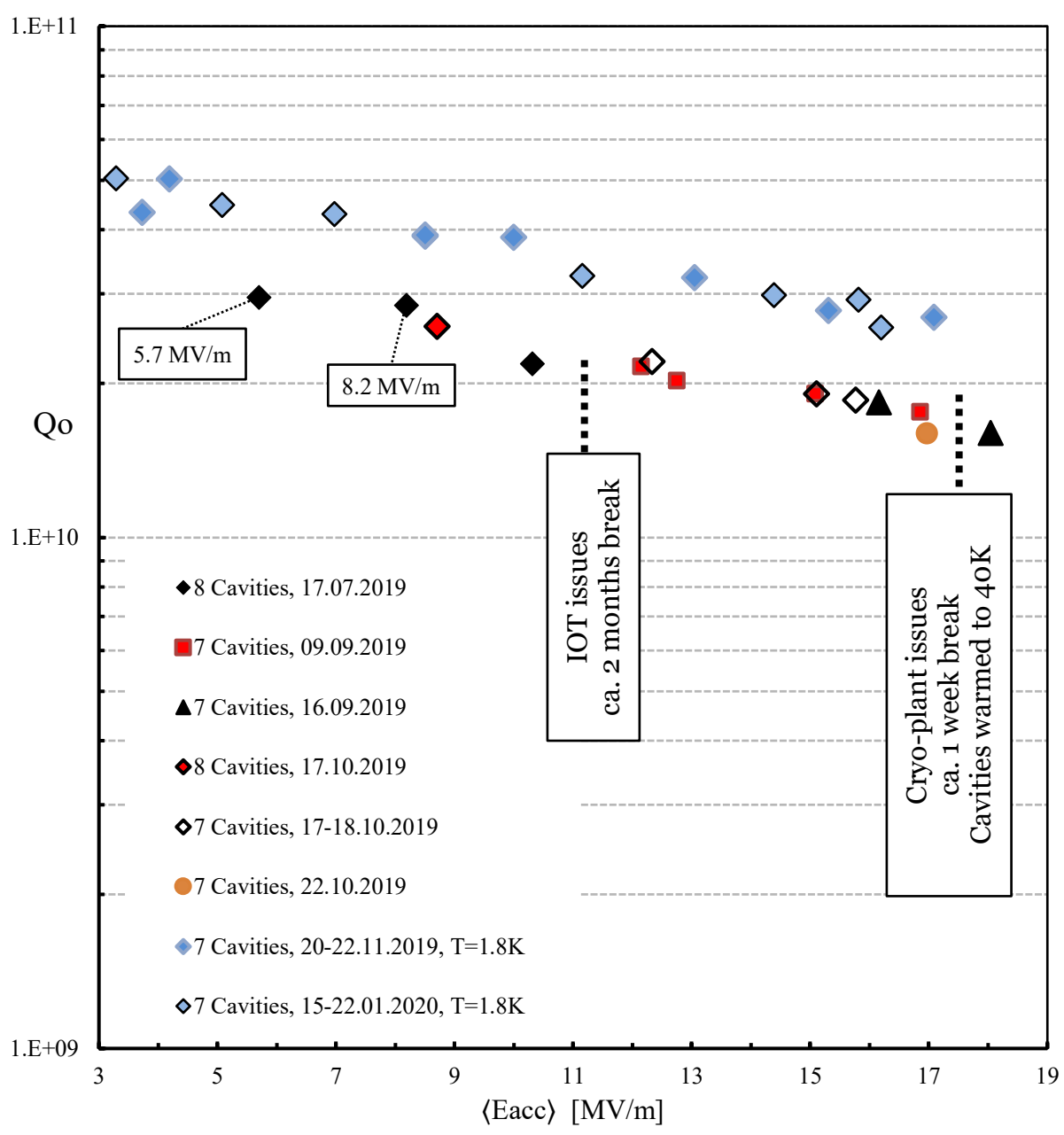

Fig. 16: Cryomodule XM50; Qo versus $\langle E_{acc}\rangle$ measured at 2 K and 1.8 K in CW mode.

Monitoring of the temperature of FPCs for seven cavities (sensors of cavity No. 4 malfunctioned from the very beginning) during the runs in July and September 2019 indicated temperature below 121 K. A summary of this tests, among the other discussed later, is presented in Figure 16.

Two days in September 2019 were devoted to the R&D programme on ANC. The test was performed on seven cavities operating at 16 MV/m in CW mode. Without ANC, the detuning range varied from ±2.5 Hz for cavity No. 8 up to approximately ±9 Hz for cavity No. 4, as shown in Figure 17. When ANC was enabled, the detuning range decreased to below ±1.5 Hz (see Figure 18). Figures 17 and 18 demonstrate the effectiveness of ANC in suppressing microphonics. This R&D topic is discussed in more detail in [12, 13].

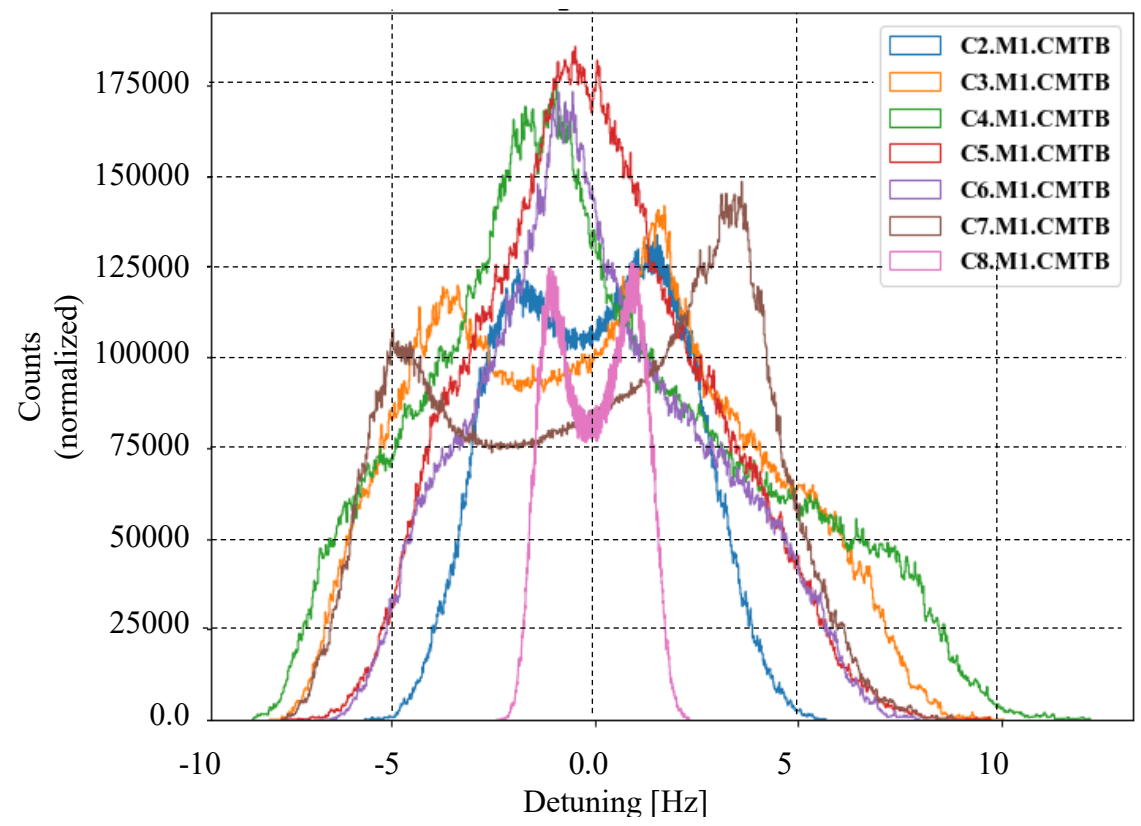

Fig. 17: Microphonics at 16 MV/m, CW mode, seven cavities, ANC off.

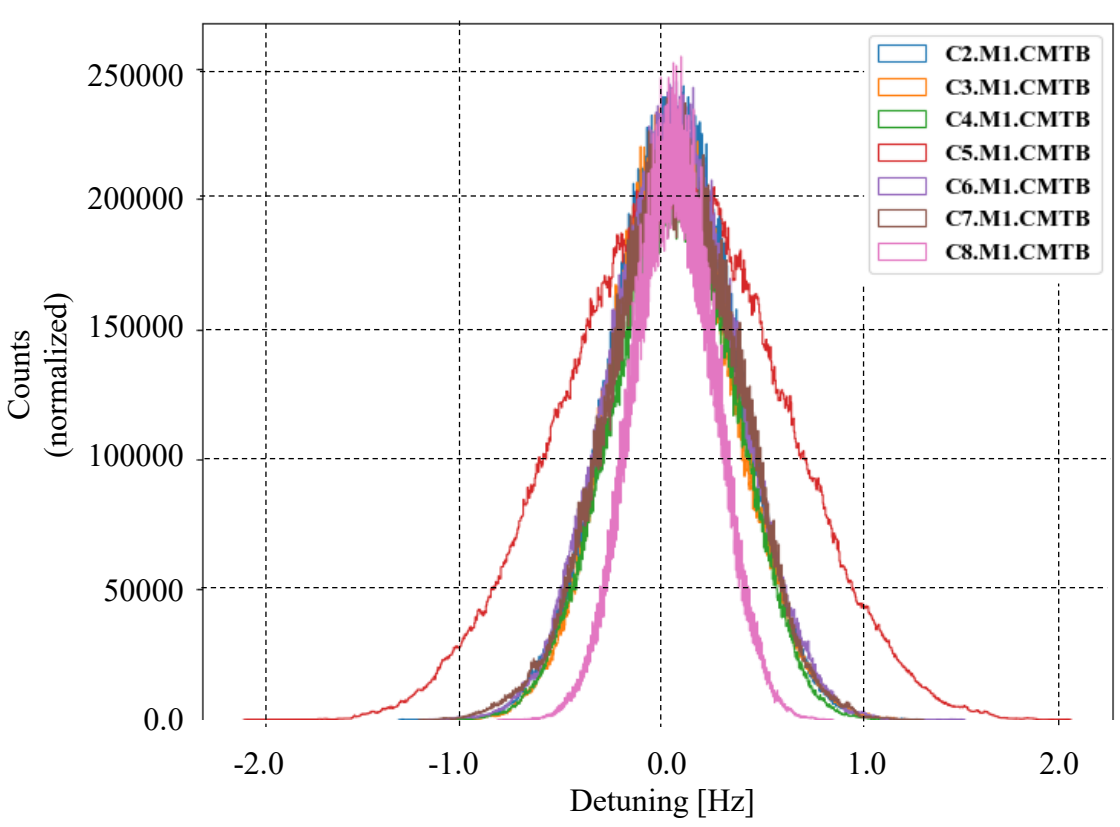

Fig. 18: Microphonics at 16 MV/m, CW mode, seven cavities, ANC on.

In the period from 17 to 22 October 2019, the inner conductors of the FPCs were retracted to their mechanical limits, resulting in the highest achievable $Q_{ext}$ values in the range of $4.1{\cdot}10^7$ to $6.1{\cdot}10^7$, and significantly lower input power was required to maintain stable operation at gradients of up to 17 MV/m. The result of the tests is shown in Figure16. For all runs in that period, the measured temperature of FPCs remained below 105 K.

Subsequently, on November 20-22, 2019 and January 15-22, 2020 we performed tests of XM50 at 1.8 K. For that tests 7 cavities were tuned on resonance. The results, marked in blue, are displayed in Figure 16. In November 2019 and on January 20, 2020, $Q_{ext}$ was set to ≈ $4{\cdot}10^7$. On January 22, 2020, $Q_{ext}$ was lowered to ≈ $3{\cdot}10^7$ due to less stable pressure in the 2 K circuit at CMTB causing gradient instability for CW operation. During these tests the temperature of FPCs was below 110 K for $Q_{ext} \approx 4{\cdot}10^7$ at $\langle E_{acc}\rangle$ = 17.1 MV/m and below 118 K $Q_{ext} \approx 3{\cdot}10^7$ at $\langle E_{acc}\rangle$ = 16.2 MV/m.

The tests of XM50 confirmed that higher $Q_{ext}$ makes possible an operation with less RF-power and thus reduces warming of the FPCs, but requires tighter other operational conditions e.g. stability of LHe pressure or less microphonics.

Several months later, on October 7th and 8th, 2020, after all cavities had been re-treated and reassembled in the cryomodule named XM50.1, the $Q_{ext}$ values of all FPCs were adjusted to the range of ≈ $4{\cdot}10^7$. The cryomodule was tested in CW mode up to an accelerating gradient of 16 MV/m for all eight cavities, and beyond this gradient with seven cavities (excluding cavity No. 2), due to instability of the cryoplant at CMTB when the DHL exceeded 130 W. The result of the test in CW mode at 2 K is shown in Figure 19. During this two-day test, we monitored the temperature of the FPCs and the power outcoupled by the HOM couplers. At an accelerating gradient of 18 MV/m, an average RF power of 2.1 kW was delivered to each FPC. The power was almost entirely reflected, as the mean power dissipated in the cavity wall was ≈ 20 W. Nonetheless, the temperature of the FPCs' remained below 126 K.

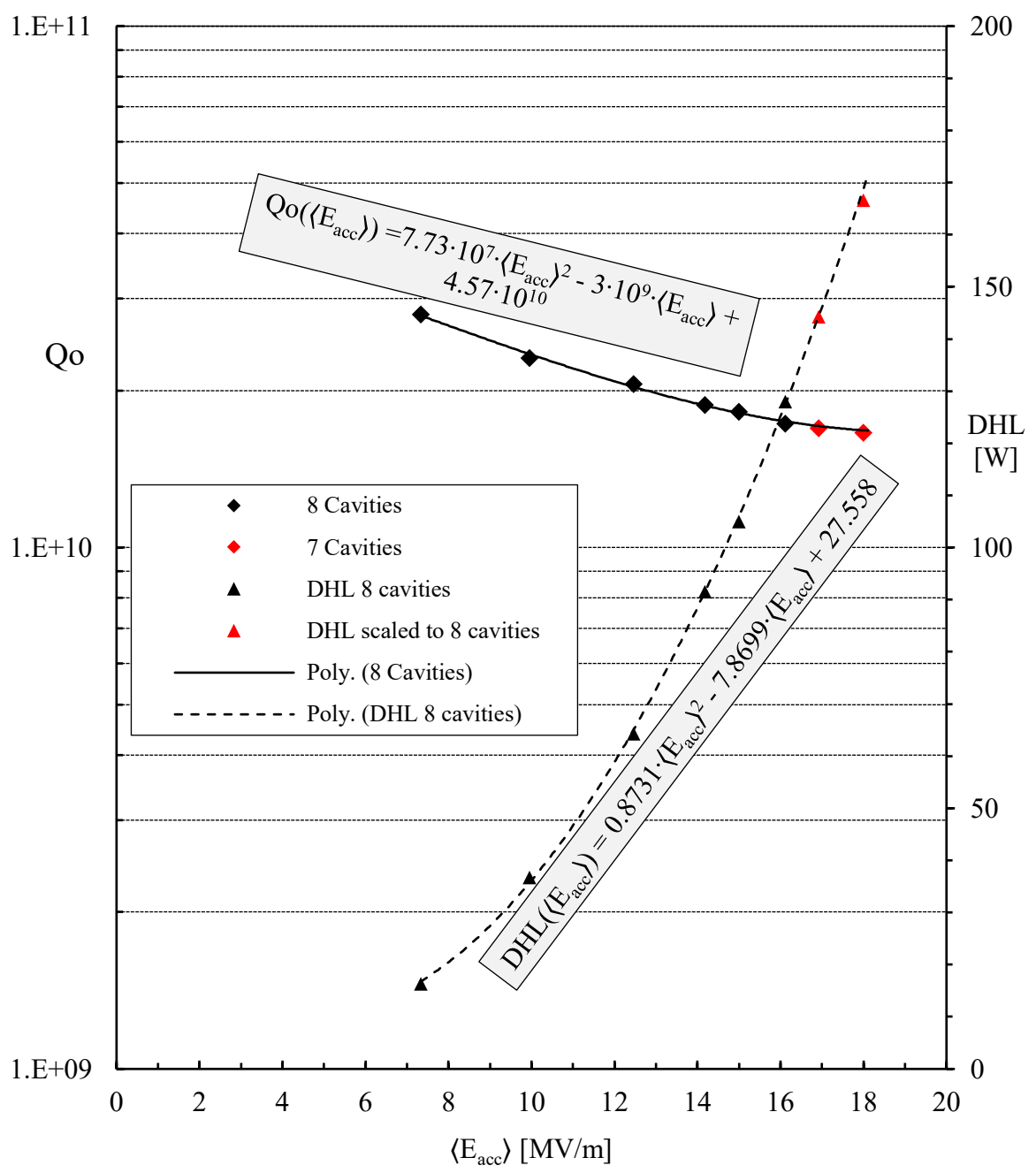

Fig. 19: Cryomodule XM50.1; $Q_0$ and DHL (secondary axis) versus $\langle E_{acc}\rangle$, measured at 2 K in CW mode. Markers indicate measured data points. The equations describe the trend lines for $Q_0$ and DHL as functions of the mean gradient $\langle E_{acc}\rangle$.

The maximum outcoupled power from the HOM couplers was ~0.35 W. The outcoupled power for all HOM couplers, at all gradients, was proportional to the stored energy in the cavities, and no changes were observed during the runs. As before, the conclusion was that the filters did not warm up. The powers outcoupled by the HOM1 and HOM2 couplers versus time, during the period in which the mean gradient was ramped up and then stabilized at $\langle E_{acc}\rangle$ = 18 MV/m, are shown in Figures 20 and 21, respectively.

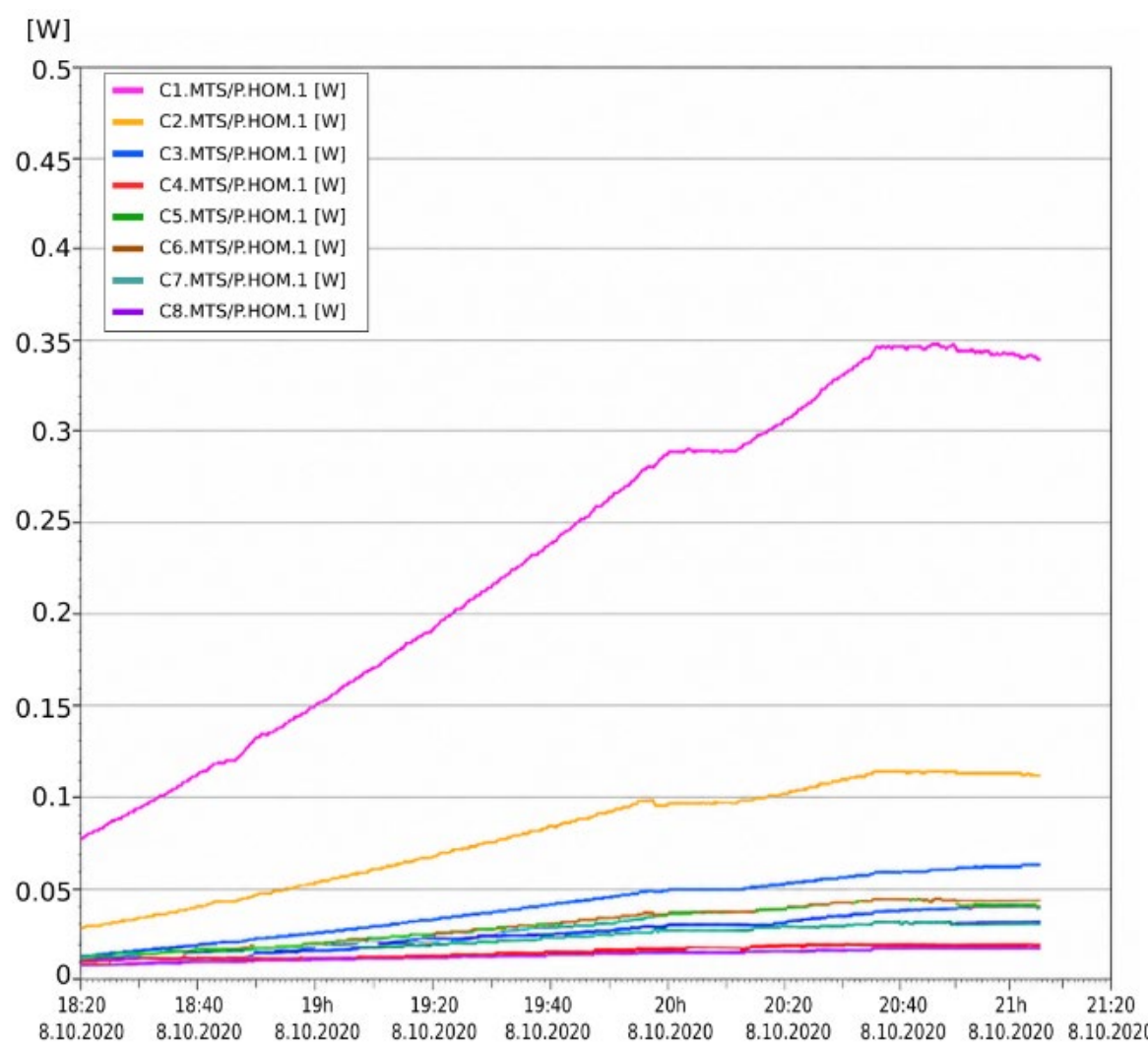

Fig. 20: The power outcoupled by the HOM1 couplers versus time during the period in which the mean gradient was ramped up to and stabilised at 18 MV/m.

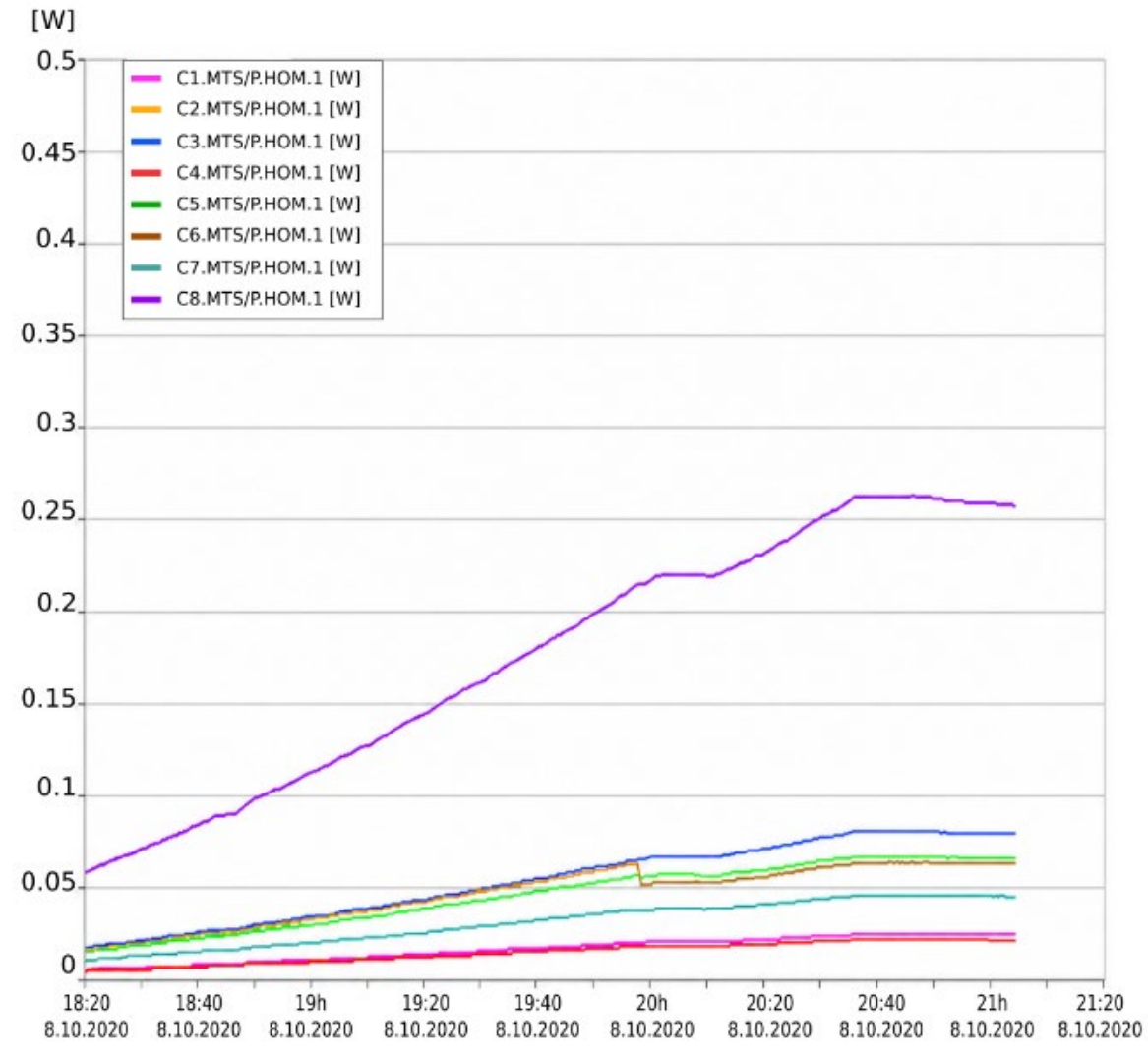

Fig. 21: The power outcoupled by the HOM2 couplers versus time during the period in which the mean gradient was ramped up to and stabilised at 18 MV/m.

In the period from May to September 2021, we tested XM46.1, the cryomodule housing re-treated cavities from the XM46 cryomodule. First, the cryomodule was tested in CW mode up to $\langle E_{acc} \rangle$ = 12 MV/m with $Q_{ext}$ of all FPCs set to $1.5 \cdot 10^7$. Then for higher gradients all FPCs the $Q_{ext}$ was set to $3 \cdot 10^7$ to reduce by half the required input power. The measured Qo and DHL as functions of $\langle E_{acc} \rangle$ are presented in Figure 22.

Next, the XM46.1 cryomodule was used to continue the LLRF R&D programme for CW and LP operation modes. The results of the LLRF R&D run are shown in Figures 23 and 24, which display the achieved amplitude and phase stability, respectively, for a mean gradient $\langle E_{acc} \rangle$ = 13 MV/m in LP operation with a duty factor of 0.55. Although the amplitude stability was very close to the specification of 0.01% for the SP operation mode, the phase stability missed the target of 0.01° by roughly a factor of five. At the time, we believed this was due to a non-optimal configuration of the ANC filters. Nevertheless, this was the first time we were able to operate with high stability in LP mode. The test is discussed in more detail in [14].

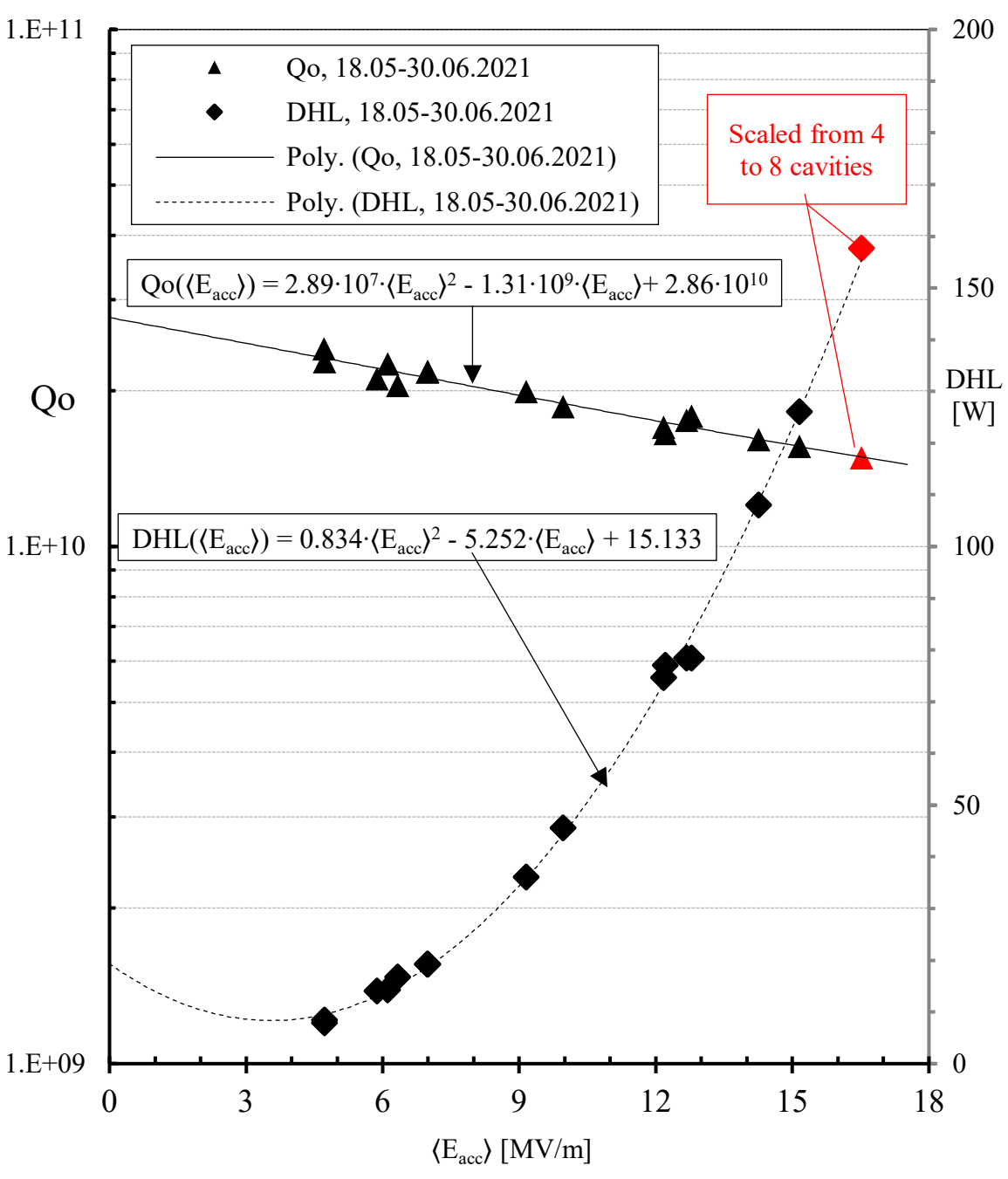

Fig. 22: Cryomodule XM46.1; $Q_0$ and DHL (secondary axis) versus $\langle E_{acc} \rangle$, measured at 2 K in CW mode. Markers indicate measured data points. The equations describe the trend lines for $Q_0$ and DHL as functions of the mean gradient $\langle E_{acc} \rangle$.

Based on the data presented in Figure 22, and assuming that the performance of XM46.1 is more representative of the cryomodules in linac L3 than the better-performing XM50.1, we estimated the number of bunches, the achievable beam on-time, and the dependence of these parameters on beam energy for CW and LP operation.

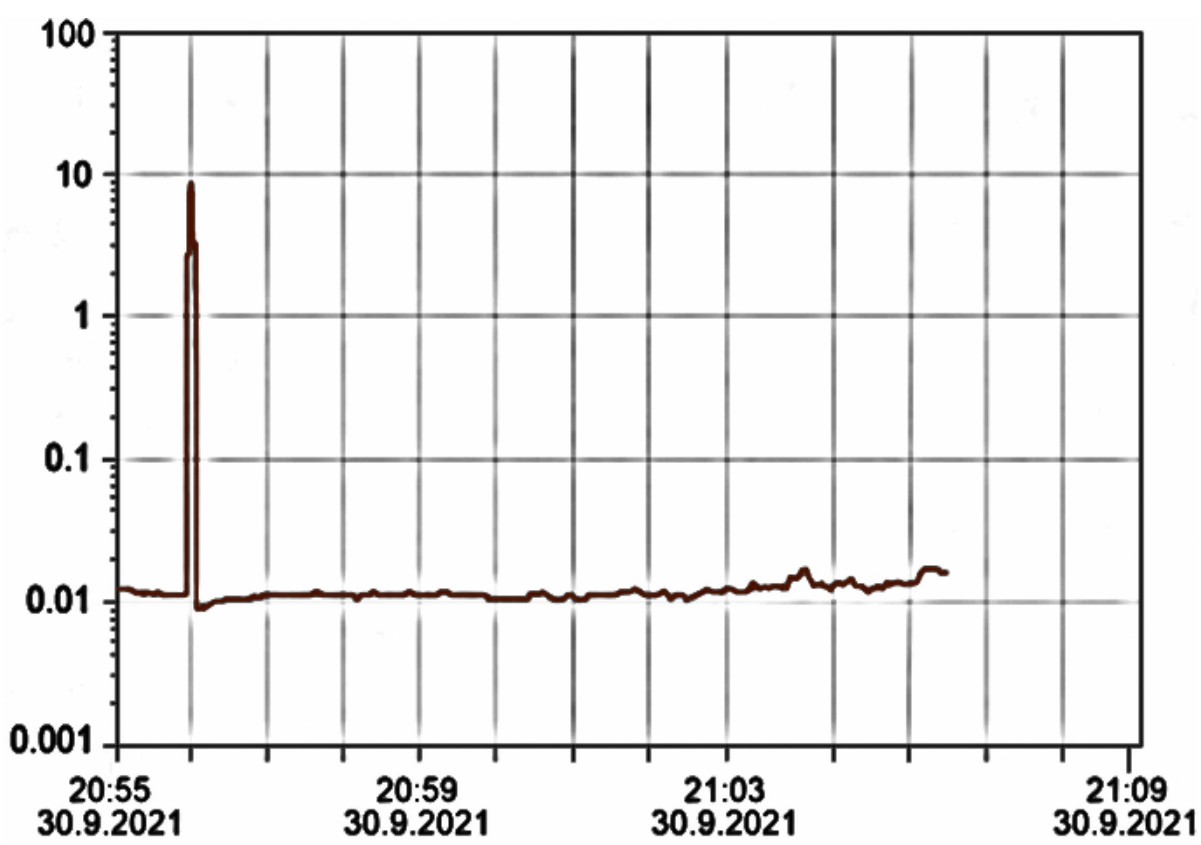

Fig. 23: LLRF system vector sum amplitude regulation performance, $\Delta\langle E_{acc} \rangle / \langle E_{acc} \rangle$ in [%].

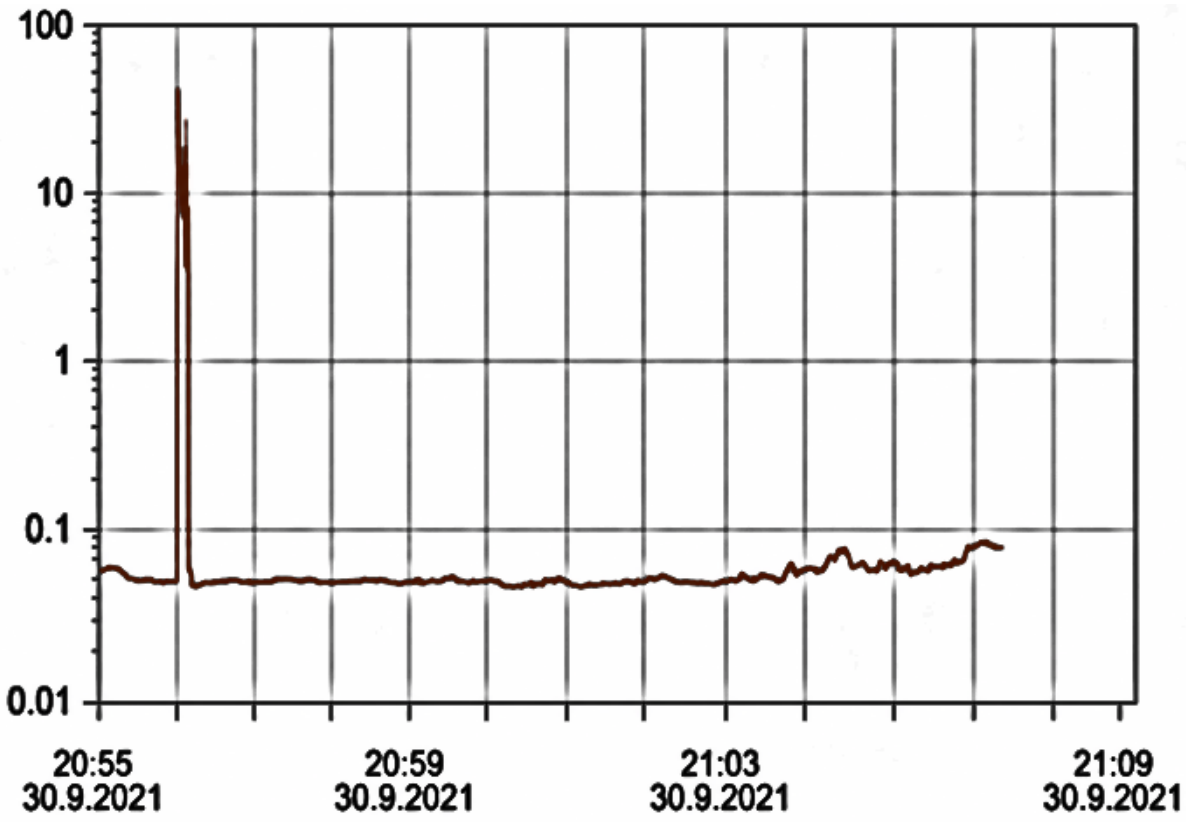

Fig. 24: LLRF system vector sum phase regulation performance, ΔP in [deg.].

The following assumptions were made for this estimation:

- The injector section delivers a 2 GeV beam to the L3 linac.
- There is a cryogenic limit, of 20 W, on the total heat load per cryomodule for the 2 K circuit.
- The 20 W is assumed to be split into 5 W static load and 15 W DHL.
- It is assumed, arbitrarily, that an electron source can deliver 250 000 equally spaced bunches per second.
- It is assumed that the RF pulse rise and decay times contribute to the cryogenic duty factor, acting with one-third of their combined duration at a gradient amplitude $\langle E_{acc} \rangle$ equal to that of the flattop. One-third of the combined duration is taken to be 17 ms.

The results based on these assumptions are presented in Figure 25. The dashed line connects all points corresponding to an RF pulse repetition rate of 1 Hz. Two additional points at 15.6 GV and 17.0 GV represent the case of a repetition rate of 0.5 Hz. The rationale behind these two LP operation points is that, when the combined duration of the rise and decay times constitutes a significant fraction of the RF pulse duration; it is advantageous to cluster two short RF pulses into a single pulse. This combined pulse has one rise and one decay phase, and a flattop longer than twice that of an individual pulse. Under cryogenic limitations, such a pulse can be initiated every other second. This approach also offers an economic advantage, as the energy stored in the entire linac is discharged half as often.

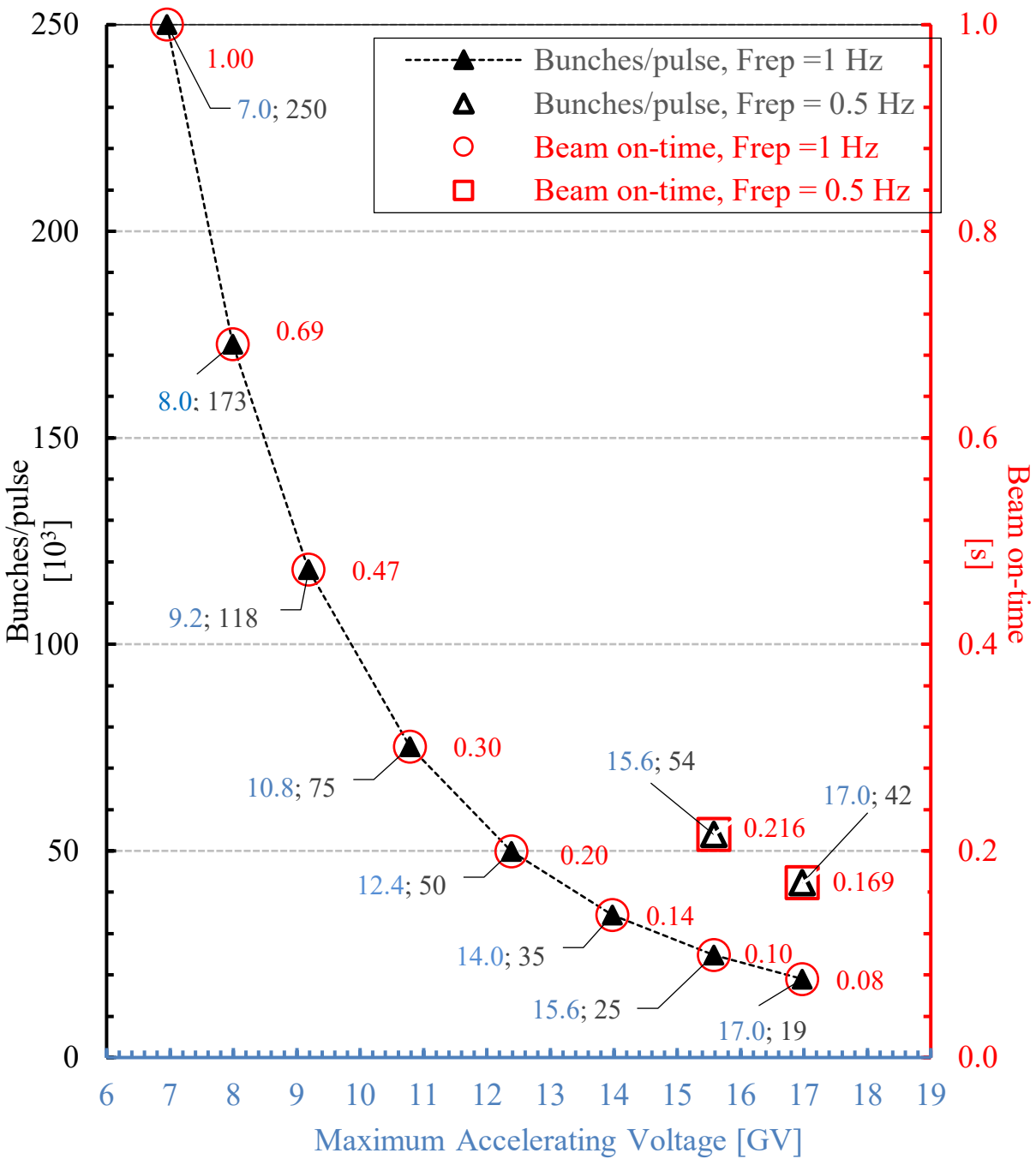

Fig. 25: Number of bunches (triangles, primary axis) and beam on-time (circles and squares, secondary axis) as a function of the maximum accelerating voltage (bunches on crest). Blue labels indicate the maximum voltage in GV, black labels indicate the number of bunches per pulse in units of $10^3$, and red labels indicate the flattop duration (beam on-time).

In September 2023, all cavities were tested individually to determine the maximum achievable gradient as a function of duty factor. Table 3 summarises the results. The last column shows the mean gradient achievable for the entire cryomodule, assuming that the RF power supplied to each cavity is sufficient to reach its maximum gradient ($E_{max}$).

Table 3: Maximum gradient versus cryogenic DF.

| | C#1 | C#2 | C#3 | C#4 | C#5 | C#6 | C#7 | C#8 | |
|---|---|---|---|---|---|---|---|---|---|
| DF [%] | Emax [MV/m] | Emax [MV/m] | Emax [MV/m] | Emax [MV/m] | Emax [MV/m] | Emax [MV/m] | Emax [MV/m] | Emax [MV/m] | <Emax> [MV/m] |
| 25 | 27.50 | 32.3 | 21.3 | 20.7 | 32.6 | 30.8 | 28.2 | 26.6 | 27.5 |
| 50 | 21.20 | 30.2 | 19.5 | 19.7 | 28.5 | 25.8 | 27.2 | 27.2 | 25 |
| 100 cw | 12.20 | 22.6 | 17.3 | 17.2 | 21.1 | 18.4 | 21.4 | 20.9 | 18.9 |

## IV. SC Photoinjector

The DESY superconducting electron source for the LP/CW upgrade was recently presented in [15, 16]. This R&D programme has continued for more than two decades. Several prototypes of the proposed injector cavity, shown in Figure 26, have demonstrated encouraging performance in a number of vertical cryogenic tests, both with directly attached superconducting Nb cathodes and with copper cathodes. Figure 27 shows data for two prototypes, 16G09 and 16G10, measured with copper cathodes.

The two new prototypes, 16G11 and 16G12, presently being fabricated in industry, will have a slightly modified shape in the vicinity of the cathode, significantly reducing electric field enhancement and thus minimising the probability of dark current generation [15]. The programme will continue at DESY, focusing primarily on establishing procedures to enhance the quantum efficiency of metallic cathodes.

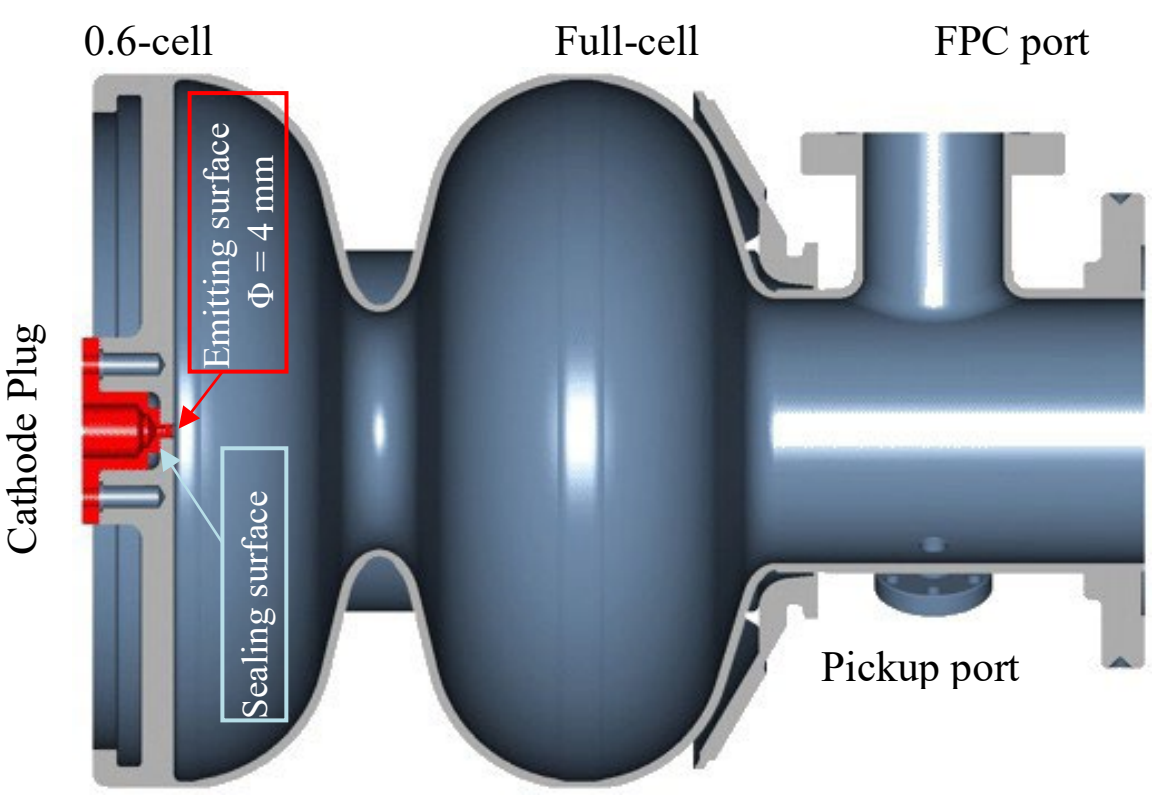

Fig. 26: DESY 1.6-cell photoinjector cavity with a cathode plug (red) attached to the backplate of 0.6-cell.

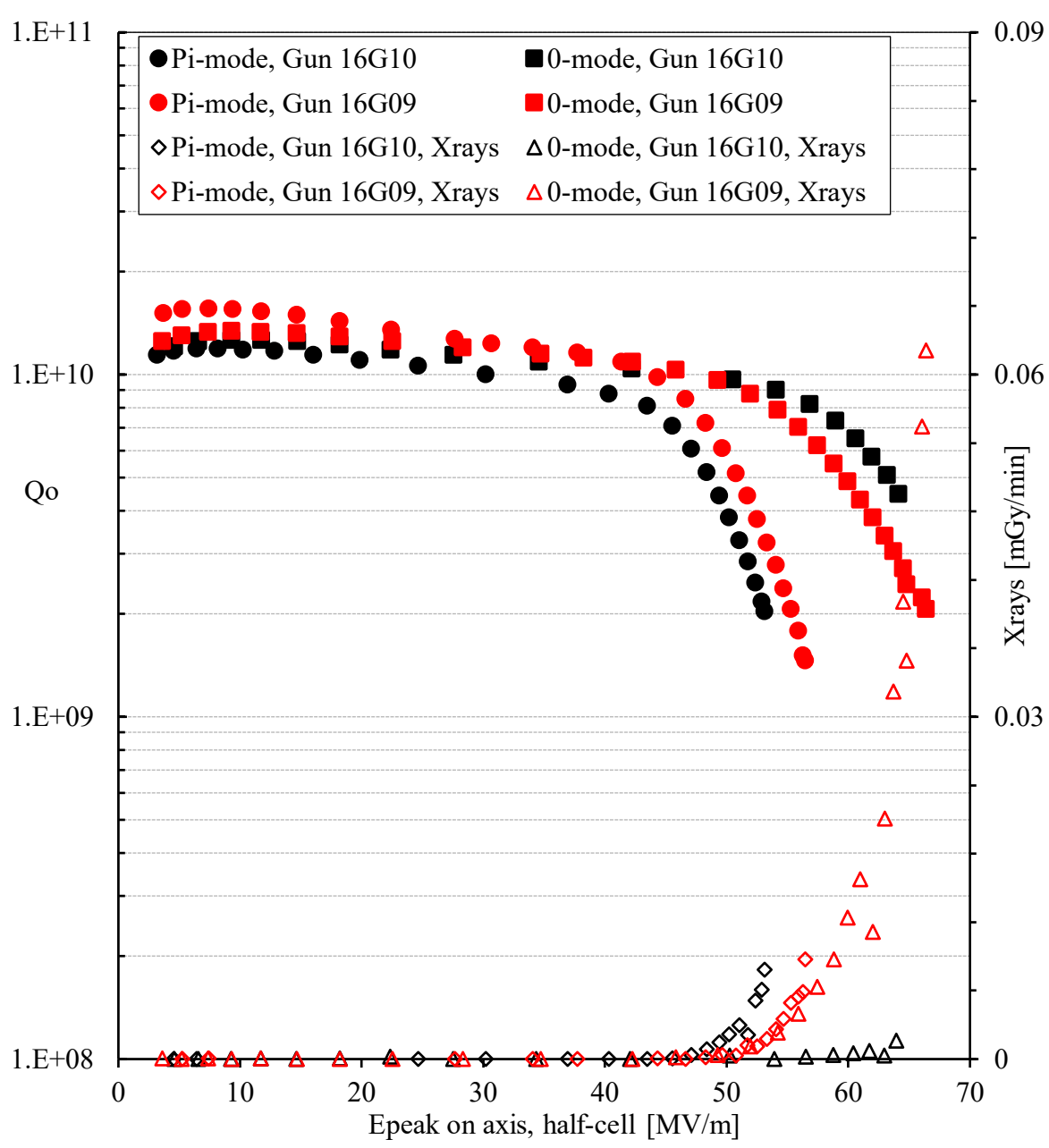

Fig. 27: Vertical test results for prototypes 16G09 and 16G10 with attached copper cathodes. The $Q_0$ data versus the peak electric field on axis in the half-cell (close to the cathode) correspond to the left-hand axis, while the X-ray data correspond to the right-hand axis.

In parallel, two injector cryomodules based on this type of cavity are currently under assembly at RI (Research Instruments GmbH) and will be delivered this year to the National Centre for Nuclear Research (NCBJ) in Poland for implementation in the Ultrafast Electron Diffraction and THz facilities. Figure 28 shows the results of vertical tests at DESY for three cavities in which the proposed shape modification has been implemented and the cathodes were made of polycrystalline niobium. The RI cavities were tested on the same test stand as all DESY prototypes.

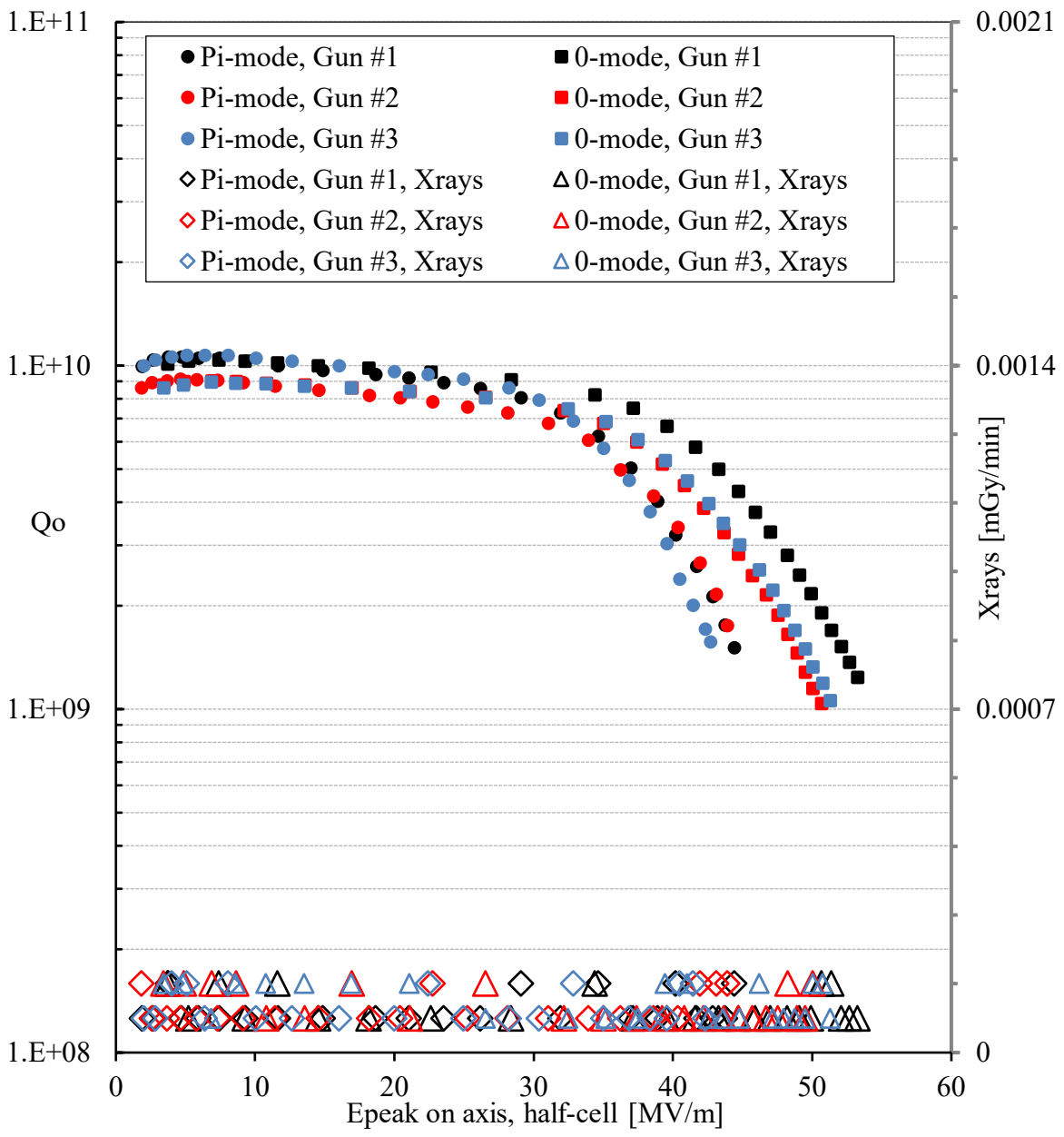

Fig. 28: Vertical test results of three RI gun cavities with attached Nb cathodes built for the NCBJ. The $Q_0$ data as a function of the peak electric field on axis in the half-cell (close to the cathode) correspond to the left-hand axis, while the X-ray data correspond to the right-hand axis.

Two conclusions can be drawn from the data presented in Figures 27 and 28.

First, the heat treatment at 95 °C for 24 hours, implemented at DESY, increases the maximum gradients by approximately 10 MV/m for both modes and results in a higher Qo. This treatment has not been applied to the RI cavities.

Second, the proposed shape modification, which reduces the enhancement of the electric field in the vicinity of the cathode from 38% to 7.5%, together with the improved RI HPR (High Pressure Rinsing) procedure, has a significant impact on the dark current. For the RI cavities, up to 44 MV/m in Pi- mode and 53 MV/m in 0-mode, no X-rays were observed, whereas for the DESY cavities with the previous shape, the onset of X-rays occurred at these gradients. This demonstrates that, for facilities operating with very low-charge bunches, such as UED, and for facilities whose electron sources must operate at high gradients, the new shape appears to be superior.

## V. Summary

From the R&D programmes and discussions on the LP/CW upgrade of the E-XFEL accelerator, which have already been ongoing for many years, the following activities will continue in the future:

a. Investigation of which type of RF power source is most suitable for the upgrade.
b. Optimization and functionality extensions implemented in the LLRF systems enabled CW and LP operation within the regulation limits, both for single-cavity control and for vector-sum control of up to eight cavities in a module. It has been demonstrated that the LLRF dedicated hardware platform, originally designed for SP operation, can be adapted to CW and/or LP modes solely through modifications of the firmware and software layers.
c. Study of quantum efficiency improvement for metallic cathodes, or the search for alternative emitters.
d. Study and development of technological procedures to enhance the $Q_0$ of the new 136 cavities (seventeen cryomodules) for the injector section.
e. Study and optimisation of the additional cryogenic plant, most probably supplying helium to the injector section.

Three of the listed R&D programmes (a, d, and e) will need to deliver final conclusions before the cost estimation phase and the start of the construction phase. Programmes b and c must provide solutions that meet the required beam parameters, such as low normalised emittance, bunch-to-bunch end-energy stability, electron energy spread, bunch length stability, and bunch charge stability. However, extending these programmes may lead to continuous improvements in the accelerator’s performance.

**Acknowledgments**

The authors wish to express their sincere gratitude to the DESY teams for their long-standing support of all the R&D programmes presented in this publication, as well as to colleagues and collaborators from BNL, HZB, HZDR, NCBJ, SLAC, and TJNAF who have contributed to advancing these activities.

The authors also wish to thank Nicolas J. Walker of DESY for motivating us to write this summary and for his assistance in editing this publication.